\documentclass{iopart} 
\usepackage{iopams}
\usepackage[colorlinks=true,citecolor=blue]{hyperref}
\usepackage{bm,amssymb,setstack}
\usepackage{graphicx,caption,subcaption}

\newcommand{\eqref}[1]{{(\ref{#1})}}    

\newcommand{ \bx }{ \mathbf{x} }

\newcommand{\bA }{ \mathbf{A} }
\newcommand{\bB }{ \mathbf{B} }
\newcommand{\bH }{ \mathbf{H} }
\newcommand{\bR }{ \mathbf{r} }
\newcommand{\bg }{ \bm{\gamma} }

\begin{document}
\title{Strong lensing, plane gravitational waves and transient flashes}
\author{Abraham I. Harte}
\address{Max-Planck-Institut f\"{u}r Gravitationsphysik, Albert-Einstein-Institut
	\\
	 Am M\"{u}hlenberg 1, D-14476 Golm, Germany}
\ead{harte@aei.mpg.de}	

\vskip 2pc

\begin{abstract}
	Plane-symmetric gravitational waves are considered as gravitational lenses. Numbers of images, frequency shifts, mutual angles, and image distortion parameters are computed exactly in essentially all non-singular plane wave spacetimes. For a fixed observation event in a particular plane wave spacetime, the number of images is found to be the same for almost every source. This number can be any positive integer, including infinity. Wavepackets of finite width are discussed in detail as well as waves which maintain a constant amplitude for all time. Short wavepackets are found to generically produce up to two images of each source which appear (separately) only some time after the wave has passed. They are initially infinitely bright, infinitely blueshifted images of the infinitely distant past. Later, these images become dim and acquire a rapidly-increasing redshift. For sufficiently weak wavepackets, one such ``flash'' almost always exists. The appearance of a second flash requires that the Ricci tensor inside the wave exceed a certain threshold. This might occur if a gravitational plane wave is sourced by, e.g., a sufficiently strong electromagnetic plane wave. 
\end{abstract}

\section{Introduction}

The theory of gravitational lensing has by now reached a considerable degree of sophistication \cite{PerlickReview, NewtonianBook, EhlersBook, PettersReview}. Theorems have been found predicting (or bounding) the number of images in very general systems \cite{OddNumber, PerlickMultImage, PerlickOddNumber}. Shapes of stable caustics have been exhaustively classified \cite{NewtonianBook,EhlersBook,CausticShapes,CausticShapes2}, a non-perturbative notion of the lens map has been obtained \cite{LensMap}, and various universal behaviors of images have been found for sources lying near caustics \cite{EhlersShape, MagInvariant, MagInvariant2}. These general results have been complemented by a number of detailed calculations for specific types of lenses. The majority of such calculations have been performed within the quasi-Newtonian viewpoint of gravitational lensing commonly used in astrophysics \cite{EhlersBook}. While various flavors of this formalism exist, most require that bending angles be small and that all lenses be nearly-Newtonian mass distributions.

This is to be contrasted with the more fundamental picture of gravitational lensing where light rays are modelled as null geodesics in a Lorentzian spacetime. Within this context, specific lensing calculations have been performed in Kerr, Reissner-Nordstr\"{o}m, and a handful of other geometries \cite{PerlickReview}. While curvatures and bending angles may be large in these examples, they all involve (at least conformally) stationary spacetimes. It is of interest to understand if qualitatively new effects appear in dynamical cases. 

This paper considers the bending of light by (decidedly non-stationary) plane-symmetric gravitational waves. Gravitational lensing by gravitational waves has previously been considered by a number of authors, although almost all of this work has been carried out within the weak-field regime \cite{DamourLensing,Kovner, EffectiveMisalignment, FaraoniRotation, FaraoniMult}. One exception is \cite{FaraoniRedshift}, where redshifts were computed in an exact solution to the vacuum Einstein equation representing a plane gravitational wave. This work considered only very specific waveforms, and was confined to a coordinate patch too small to include caustics and other effects associated with the formation of multiple images. Separately, extensive work has been devoted to non-perturbatively understanding the geodesic structure of generic plane wave spacetimes \cite{EhrlichTameWave, Ehrlich2, GlobalGeometry, HarteCaustics, Hollowood1, Hollowood2, Bondi2}. This is clearly a subject closely related to gravitational lensing, although few explicit relations between the two subjects appear to have been made (see, however, remarks in \cite{PerlickReview}). 

It is the purpose of this work to provide a comprehensive and non-perturbative discussion of lensing in plane wave spacetimes. These geometries are a well-known subclass of \textit{pp}-waves; plane-fronted waves with parallel rays. Many plane wave spacetimes are exact solutions to the vacuum Einstein equation. Others may be interpreted as, e.g., exact solutions in Einstein-Maxwell theory. While rather idealized from the astrophysical perspective, plane wave spacetimes admit a wide variety of interesting phenomena. Depending on the waveform, any number of astigmatic and anastigmatic caustics may exist. Examples admitting any specified number of images -- even an infinite number -- are easily constructed. In particular, even numbers of images can exist \cite{PerlickReview}. The image count can also change in time even when a source does not cross an observer's caustic. Despite all of these properties, plane wave spacetimes are geometrically very mild. They are topologically equivalent to $\mathbb{R}^4$ and admit coordinate systems which cover the entire manifold.

Aside from their value as models of gravitational radiation, the plane wave spacetimes considered here have also found numerous applications via the Penrose limit. This limit provides a sense in which the metric near any null geodesic in any spacetime is equivalent to the metric of an appropriate plane wave spacetime \cite{PenroseLimit, BlauPenrose}. It allows problems in relatively complicated spacetimes to be reduced to equivalent problems in plane wave spacetimes (which are often much simpler). This has been particularly valuable within string theory and related fields \cite{Maldacena, String2}. Penrose limits have also been applied to ordinary quantum field theory in order to investigate causality and effective indices of refraction for photons and gravitons propagating in curved spacetimes \cite{Hollowood1, Hollowood2, Hollowood3}. More recently, Penrose limits were used to deduce the effect of caustics on Green functions associated with the propagation of classical fields in curved spacetimes \cite{HarteCaustics}. Given the content of the Penrose limit, lensing in plane wave spacetimes might be related to lensing in generic spacetimes as seen by ultrarelativistic observers. We make no attempt to justify this conjecture, however.

This paper starts by providing a self-contained review of plane wave spacetimes in Sect. \ref{Sect:PlaneReview}. Although most of this material is not new \cite{EhrlichTameWave, Ehrlich2, GlobalGeometry, HarteCaustics, Hollowood1, Hollowood2, Bondi2, BondiPW}, it is not widely known. Sect. \ref{Sect:NumImages} then derives the number of images of a point source that may be viewed in plane wave spacetimes. Under generic conditions, this is found to depend only on the waveform and a certain time parameter associated with the observation event. The number of images does not depend on any properties of the source. Once this is established, Sect. \ref{Sect:ImageProperties} computes image positions, frequency shifts, angles, and image distortion parameters in general plane wave spacetimes. Sect. \ref{Sect:SymWaves} applies these results to symmetric plane waves, which have constant waveforms. These geometries produce an infinite number of images of almost every source. Their lensing properties are found to change significantly if the Ricci tensor is increased beyond a certain threshold. Lastly, Sect. \ref{Sect:Sandwaves} discusses ``sandwich waves;'' wavepackets with finite width. These spacetimes generically admit images which appear at discrete times and then persist indefinitely. Such images initially provide infinitely blueshifted, infinitely bright pictures into the infinitely distant past. Very quickly, however, such images become highly redshifted and effectively fade away. One of these ``transient flashes'' is produced by almost every sufficiently weak vacuum (Ricci-flat) wave. For weak waves, a second flash appears only if the Ricci tensor of the wavepacket exceeds a certain threshold. Throughout this work, the spacetime is assumed to be everywhere transparent. The language used also assumes that the geometric optics approximation \cite{EhlersBook} holds even in situations where it would be severely strained (such as when light is emitted near an observer's caustic).

\subsection*{Notation}

This paper restricts attention only to plane wave spacetimes in four spacetime dimensions. Our sign conventions follow those of Wald \cite{Wald}. The signature is $-+++$. Latin letters $a, b, \ldots$ (and occasionally $A, B, \ldots$) from the beginning of the alphabet are used to denote abstract indices. Greek letters $\mu, \nu, \ldots$ are used to denote four-dimensional coordinate indices. The Latin letters $i, j, \ldots$ are instead coordinate indices associated with the two directions transverse to the direction of wave propagation. Objects involving the latter type of index are often written in boldface with all indices suppressed. They are then manipulated using the standard notation of linear algebra [e.g., $A_{ki} B_{kj} = ( \bA^\intercal \bB )_{ij}$ and $|\bx| = \sqrt{ x_i x_i }$]. Overall, notation related to plane wave spacetimes closely follows the conventions of \cite{HarteCaustics}.

\section{Geometry of plane wave spacetimes}
\label{Sect:PlaneReview}

Plane wave geometries may be interpreted as simple models for gravitational waves emitted from distant sources. Alternatively, they arise as universal limits for the geometries near null geodesics in any spacetime \cite{PenroseLimit,BlauPenrose}. The typical definition of a plane wave spacetime $(M,g_{ab})$ requires that $M = \mathbb{R}^4$ and that there exist global coordinates $(u,v,\bx) = (u,v,x^1, x^2): M \rightarrow \mathbb{R}^4$ such that the line element takes the form
\begin{equation} 
	g_{\mu\nu} \rmd x^\mu \rmd x^\nu= - 2 \rmd u \rmd v + H_{ij}(u) x^i x^j \rmd u^2 +  (\rmd x^1)^2 + (\rmd x^2)^2 .
	\label{PlaneWaveMetric}
\end{equation}
$H_{ij} = (\mathbf{H})_{ij}$ is any symmetric $2 \times 2$ matrix. Its components describe the waveforms associated with a wave's three polarization states\footnote{One of the three polarization states associated with plane wave spacetimes vanishes in the vacuum case $R_{ab}=0$. This leaves the usual two polarizations associated with vacuum general relativity. Note that the six polarization states typically stated to exist for linearized ``plane'' gravitational waves in generic theories of gravity \cite{GWPolarizations} cannot all be represented by the metric \eqref{PlaneWaveMetric}. Three of these polarizations may be realized only as geometries which are rather less plane-symmetric than those considered here.}. The $u$ coordinate is interpreted as a phase parameter for the wave, while $v$ affinely parametrizes its rays. The remaining two coordinates $x^i$ span spacelike wavefronts transverse to the wave's direction of propagation.

Note that if $\bH =0$ in some region, the spacetime is locally flat there. In terms of a Minkowski coordinate system $(t,x^1, x^2, x^3)$, $u$ and $v$ satisfy $u=(t+x^3)/\sqrt{2}$ and $v=(t-x^3)/\sqrt{2}$ in such a region. We consider only nontrivial plane waves, so $\bH$ cannot vanish everywhere.

The physical interpretation of the so-called Brinkmann metric \eqref{PlaneWaveMetric} as a plane-symmetric gravitational wave follows from considering the integral curves of the vector field $\ell^a = (\partial/ \partial v)^a$. These curves form a null geodesic congruence which may be interpreted as the rays of the gravitational wave. $\nabla_a \ell^b=0$, so these rays have vanishing expansion, shear, and twist. There is therefore a sense in which they are everywhere parallel to one another. All rays are also orthogonal to the family of spacelike 2-surfaces generated by the two commuting spacelike vector fields $X_{(i)}^a = (\partial / \partial x^i)^a$. The induced metric on each such surface is flat: The wavefronts are 2-planes. The curvature is constant on these planes in the sense that the $X^a_{(i)}$ are curvature collineations: 
\begin{equation}
	\mathcal{L}_{X_{(i)}} R_{abc}{}^{d} = 0.
\end{equation}
Despite this, the $X^a_{(i)}$ are not everywhere Killing. There do, however, exist linear combinations of $X_{(i)}^a$ and $\ell^a$ which \textit{are} Killing. $\ell^a$ itself is also Killing, which may be interpreted as a statement that plane waves do not deform along their characteristics. Plane wave spacetimes admit a minimum of five linearly independent Killing fields. Note that in flat spacetime, five (out of the total of ten) Killing fields are symmetries of all electromagnetic plane waves \cite{BondiPW}. Killing fields of plane wave spacetimes are discussed more fully in Sect. \ref{Sect:Geodesic}.

All non-vanishing coordinate components of the Riemann tensor may be determined from
\begin{equation}
	R_{uiuj} = - H_{ij}.
\label{Riemann}
\end{equation}
It follows from this that the Ricci tensor is
\begin{equation}
	R_{ab} = - \Tr  \mathbf{H} \ell_a \ell_b,
\label{Ricci}
\end{equation}
where  $\Tr$ denotes the ordinary (Euclidean) trace of the $2 \times 2$ matrix $\mathbf{H}$. The Ricci scalar always vanishes in plane wave spacetimes. More generally, there are no nonzero scalars formed by local contractions of the metric, the curvature, and its derivatives: $R^{ab} R_{ab} = R_{abcd} R^{abcd}  =  R_{abcd}  \epsilon^{abfh} R_{fh}{}^{cd} = \ldots = 0$. This is analogous to the fact that plane electromagnetic waves satisfy, e.g., $F_{ab} F^{ab} = \epsilon^{abcd} F_{ab} F_{cd} = 0$. Note, however, that electromagnetic plane waves are not the only electromagnetic fields with vanishing field scalars. Similarly, plane wave spacetimes are not the only curved geometries with vanishing curvature scalars \cite{VSIspacetimes}.

It follows from \eqref{Ricci} that plane wave spacetimes satisfying the vacuum Einstein equation (and the vacuum equations of many alternative theories of gravity \cite{ModGravity}) are characterized by the simple algebraic constraint $\Tr \mathbf{H}= 0$. For vacuum waves, there exist two scalar functions $h_{+}$ and $h_\times $ such that 
\begin{equation}
	\mathbf{H} = \left( \begin{array}{cc}
									-h_{+} & h_\times \\
									h_\times & h_{+}
							\end{array} \right).
	\label{HVac}
\end{equation}
$h_+$ and $h_\times$ describe the waveforms for the two polarization states of a gravitational plane wave propagating in vacuum. A plane wave is said to be linearly polarized if $h_+$ and $h_\times$ are linearly dependent (in which case one of these functions can be eliminated by a suitable rotation of the transverse coordinates $\bx$). 

If $h_+$ and $h_\times$ have compact support, the geometry is said to be a sandwich wave. This name evinces the image of a curved region of spacetime ``sandwiched'' between null hyperplanes in a geometry that is otherwise Minkowski. Physically, it corresponds to a wavepacket of finite length. Note that the planar symmetry considered here is very special in the sense that passing waves do not necessarily leave any ``tail'' behind them. After interacting with a sandwich wave, all observers enter a region of spacetime which is perfectly flat. There is a sense in which test fields propagating on plane wave spacetimes also have no tails \cite{HarteCaustics, HuygensBook}. 

A general (not necessarily vacuum) wave profile $\mathbf{H}$ may be built by adding to \eqref{HVac} a term proportional to the identity matrix $\bm{\delta}$. There then exists a third polarization function $h_\|$ such that
\begin{equation}
	\mathbf{H} = \left( \begin{array}{cc}
							-h_{+} - h_{\|}  & h_\times \\
							h_\times & h_{+} - h_{\|}
						\end{array} \right).
						\label{HGen}
\end{equation}
If the Ricci tensor of such a wave is associated with a stress-energy tensor via Einstein's equation, that stress-energy tensor obeys the weak energy condition if and only if $h_{\|} \geq 0$. Assuming this, the stress-energy tensors associated with \eqref{HGen} are very simple. They could be generated by, e.g., electromagnetic plane waves with the form
\begin{equation}
	F_{ab} = 2 h^{ \frac{1}{2} }_{\|}  \nabla_{[a} u \nabla_{b]} x^1.
	\label{EMField}
\end{equation}
Alternatively, \eqref{HGen} could be associated with the stress-energy tensor of the massless Klein-Gordon plane wave
\begin{equation}
	\phi = \int_{u}  h^{ \frac{1}{2} }_\| (w) \rmd w.
	\label{ScalarField}
\end{equation}

Besides the vacuum case $h_\| = 0$, another interesting class of wave profiles are those that are conformally flat. These satisfy $h_+ = h_\times = 0$, so $\mathbf{H} \propto \bm{\delta}$. As gravitational lenses, all caustics of conformally-flat plane waves are associated with ``perfect'' anastigmatic focusing. For more general plane waves, caustics are typically (but not necessarily) associated with astigmatic focusing.

\subsection{The matrices $\bA $ and $\bB $}
\label{Sect:AB}

The geometry of plane wave spacetimes has been analyzed in detail by a number of authors \cite{EhrlichTameWave, Ehrlich2, GlobalGeometry, HarteCaustics, Hollowood1, Hollowood2, Bondi2, BondiPW}. One essential conclusion of this work is that nearly all interesting properties of plane wave spacetimes may be deduced from the properties of $2 \times 2$ matrices $\mathbf{E} = \mathbf{E}(u)$ satisfying the differential equation
\begin{equation}
	\ddot{\mathbf{E}} = \mathbf{H} \mathbf{E}.
	\label{EDef}
\end{equation}
This is a ``generalized oscillator equation'' with $-\bH $ acting like a matrix of squared frequencies. Eq. \eqref{EDef} arises when solving for geodesics or Jacobi fields in plane wave spacetimes. Bitensors such as Synge's function and the parallel propagator may be written explicitly in terms of its solutions. The same is also true for a plane wave's Killing vectors.

It is convenient to write all possible matrices $\mathbf{E}$ in terms of two particular solutions. Fix any\footnote{The notation $u_o$ is used here because this will later be interpreted as a $u$ coordinate associated with some observer. Similarly, $u_s$ is often interpreted below as a $u$ coordinate associated with a source (not necessarily when that source is observable at any observation point with $u=u_o$). Later, the notation $u_e$ is used to denote a $u$ coordinate associated with the emission of light from a source.} $u_o \in \mathbb{R}$ and define $\bA(\cdot,u_o)$ and $\bB(\cdot,u_o)$ to be solutions to \eqref{EDef} (where derivatives are applied to the first arguments of $\bA $ and $\bB $) with the initial conditions
\begin{eqnarray}
	\lim_{u_s \rightarrow u_o} \bA(u_s,u_o) = \lim_{u_s \rightarrow u_o} \partial_{(1)} \bB(u_s,u_o) = \bm{\delta}, 
\\
	\lim_{u_s \rightarrow u_o} \partial_{(1)} \bA (u_s,u_o)= \lim_{u_s \rightarrow u_o} \bB (u_s,u_o) = 0.
	\label{InitAB}
\end{eqnarray}
Here, the notation $\partial_{(1)} \bA $ indicates a partial derivative with respect to the first argument of $\bA $. We assume for simplicity that $\bH $ is a matrix of piecewise-continuous functions and that $\bA $ and $\bB$ are at least $C^1$ (and piecewise-$C^2$) in both of their arguments. Example expressions for $\bA $ and $\bB $ are discussed in Sects. \ref{Sect:SymWaves} and \ref{Sect:Sandwaves}.

Given any two solutions $\mathbf{E}_1$ and $\mathbf{E}_2$ to \eqref{EDef}, it is easily verified that their Wronskian is conserved:
\begin{equation}
	\mathbf{E}_1^\intercal \dot{\mathbf{E}}_2 - \dot{\mathbf{E}}_1^\intercal \mathbf{E}_2 = \mbox{constant}.
\label{GenWronskian}
\end{equation}
Here, ${}^\intercal$ denotes a matrix transpose. Applying this formula with $\mathbf{E}_1 \rightarrow \bA $ and $\mathbf{E}_2 \rightarrow \bB $ shows that
\begin{equation}
	\bA ^\intercal \partial_{(1)} \bB - \partial_{(1)} \bA ^\intercal \bB = \bm{\delta}.
\label{Wronskian}
\end{equation}
Neither $\bA$ nor $\bB$ are necessarily symmetric matrices. Nevertheless, \eqref{GenWronskian} and \eqref{Wronskian} may be used to show that the products 
\begin{equation}
\fl \qquad \qquad 
	\bA^\intercal \partial_{(1)} \bA, 
\quad
	\partial_{(1)} \bA \bA^{-1} 
\quad
	\bB^\intercal \partial_{(1)} \bB, 
\quad
	\partial_{(1)} \bB \bB^{-1},
\quad
	\bB \bA^\intercal,
\quad
	\bB ^{-1} \bA
\label{SymMatrices}
\end{equation}
\textit{are} symmetric wherever they exist \cite{HarteCaustics}. Letting $\mathbf{E}_1 \rightarrow \bB(\cdot,u_o)$ and $\mathbf{E}_2 \rightarrow \bB (\cdot,u_s)$ in \eqref{GenWronskian} shows that
\begin{equation}
	\bB (u_s,u_o) = - \bB^\intercal (u_o,u_s) 
	\label{BSwitch}
\end{equation}
for all $u_s, u_o \in \mathbb{R}$. This is essentially Etherington's reciprocity law \cite{PerlickReview,EhlersBook,Etherington, EllisEth}. A similar calculation may be used to show that
\begin{equation}
	\partial_{(1)} \bA (u_s,u_o) = - \partial_{(1)} \bA^\intercal (u_o,u_s)
\end{equation}
as well.

It is sometimes useful to consider partial derivatives $\partial_{(2)}$ with respect to the second arguments of $\bA $ and $\bB $. The resulting matrices remain solutions to \eqref{EDef}. Comparing initial conditions shows that 
\begin{equation}
\fl \qquad	\partial_{(2)} \bA (u_s,u_o) = - \bB(u_s,u_o) \mathbf{H}(u_o), 	\qquad
	\partial_{(2)} \bB (u_s,u_o) = - \bA (u_s,u_o) .
	\label{PrimedD}
\end{equation}
$\bA $ may therefore be derived from $\bB $. The opposite is also true wherever $\det \bH \neq 0$. 

Geometrically, $\bB$ corresponds to the transverse coordinate components of a Jacobi propagator describing the evolution of deviation vectors along geodesics passing between different pairs of points \cite{HarteCaustics}. $\bB $ is also related to image distortion. Up to an overall time dilation factor, it translates small differences in image position on an observer's sky to spatial deviations from a fiducial source point. In the language of \cite{EhlersShape}, $\bB $ is proportional to the Jacobi map. The symmetric matrix $\partial_{(1)} \bB \bB ^{-1}$ plays a similar role, but translates source separations to \textit{emission} (rather than observation) angles. It is proportional to an object typically referred to as the optical deformation matrix. These statements are explained more fully in Sect. \ref{Sect:ImageDist}.

\subsection{Conjugate pairs}
\label{Sect:Conjugate}

It is often useful when working with plane wave spacetimes to consider hypersurfaces of ``constant phase.'' Recalling the interpretation of the $u$ coordinate as a phase, let
\begin{equation}
	S_{u_o} := \{ p \in M : u(p) = u_o \}
\end{equation}
denote such a hypersurface. Given two points $p_s \in S_{u_s}$ and $p_o \in S_{u_o}$ (with $u_s \neq  u_o$), define the multiplicity or ``index'' of these points to be \cite{EhrlichTameWave}
\begin{equation}
	I(p_s,p_o) := 2 - \mathrm{rank}\, \bB (u_s,u_o). 
	\label{IDef}
\end{equation}
We also set $I(p_o,p_s) = 0$ whenever $u_o =u_s$. It follows from \eqref{BSwitch} that $I(p_s,p_o) = I(p_o, p_s)$. For most pairs of points, $I = 0$. Such pairs are said to be ``disconjugate.'' Pairs $p_s,p_o$ satisfying $I(p_s,p_o) \neq 0$ are instead said to be conjugate with multiplicity $I(p_s,p_o)$. Similarly, we call the pairs $S_{u_s}, S_{u_o}$ ``conjugate hyperplanes'' and the pairs of real numbers $u_s,u_o$ ``conjugate phases'' when $\mathrm{rank} \, \bB (u_s,u_o)<2$. Despite the appearance of \eqref{IDef}, the index map $I: M \times M \rightarrow \{ 0,1,2\}$ describes phenomena which do not depend on any choice of coordinate system.

Conjugate pairs as described here are closely related to the conjugate points commonly considered in differential geometry and optics. In general, distinct points $p_s$ and $p_o$ on a given geodesic are said to be conjugate if and only if there exist nontrivial deviation vectors along that geodesic which vanish at both $p_o$ and $p_s$. In plane wave spacetimes, this condition reduces to $I(p_s, p_o) >0$. Defining the multiplicity of a pair of conjugate points to be the number of linearly independent deviation vectors which vanish at those points, that multiplicity is equal to $I(p_s, p_o)$. The concept of conjugacy associated with $I$ does not, however, require the specification of any particular geodesic. It is uniquely defined even for pairs of points connected by multiple geodesics or by none. Indeed, these are the only cases where $I \neq 0$.

All strong lensing effects associated with plane wave spacetimes are related to the existence of conjugate hyperplanes. It follows from \eqref{IDef} that every pair of conjugate phases $u_s,u_o$ satisfies
\begin{equation}
	\det \bB (u_s, u_o) = \det \bB (u_o, u_s)  = 0.
\label{ZeroDet}
\end{equation}
Finding conjugate pairs and their multiplicities may be viewed as a matter of direct computation once $\bH $ is specified. Alternatively, various Sturm-type comparison theorems can be used to make general statements regarding the existence and separations of conjugate pairs for various classes of plane wave. See, e.g., Chapt. XI of \cite{Hartman} for results relating to mathematical problems of this type and \cite{EhrlichTameWave} for an application to ``tame'' plane wave spacetimes. More specific examples are discussed in Sects. \ref{Sect:SymWaves} and \ref{Sect:Sandwaves} below. 

The qualitative structure of geodesics in plane wave spacetimes is closely related to the index $I$. First note that every disconjugate pair of points $p_s, p_o$ is connected by exactly one geodesic. If two points are conjugate, the number of connecting geodesics is either zero or infinity. Sizes of geodesically connected regions may be summarized by
\begin{equation}
	\fl \qquad \dim [ (\mbox{all points geodesically connected to $p_o$}) \cap S_{u_s}] + I (p_s, p_o) = 3 .
\label{IndexTheorem}
\end{equation}
A similar relation exists for null cones when $u_o \neq u_s$:
\begin{equation}
	\fl \qquad \dim [ (\mbox{all points connected to $p_o$ via null geodesics}) \cap S_{u_s}] + I (p_s, p_o) = 2 .
\label{IndexTheorem2}
\end{equation}
The latter result has been referred to as an ``index theorem'' in \cite{EhrlichTameWave}. Eq. \eqref{IndexTheorem} implies that geodesics emanating from $p_o$ and intersecting a hyperplane $S_{u_s}$ that is disconjugate to $S_{u_o}$ form a 3-dimensional region. Indeed, these geodesics fill the entire hyperplane. More interestingly, geodesics intersecting a conjugate hyperplane with multiplicity 1 fill only a 2-dimensional region on that 3-dimensional surface. Geodesics intersecting a hyperplane with multiplicity 2 form a line. Similarly, the null cone of a point reduces to a 1-dimensional curve on every hyperplane with multiplicity 1. A null cone intersecting a hyperplane with multiplicity 2 is focused to a single point on that hyperplane. These two cases correspond to astigmatic and anastigmatic focusing, respectively. Anastigmatic focusing tends to be unstable in the sense that perturbations tend to split a single multiplicity 2 phase into two closely-spaced phases each with multiplicity 1.

\begin{figure}
	\centering
	\includegraphics[width=.45\linewidth]{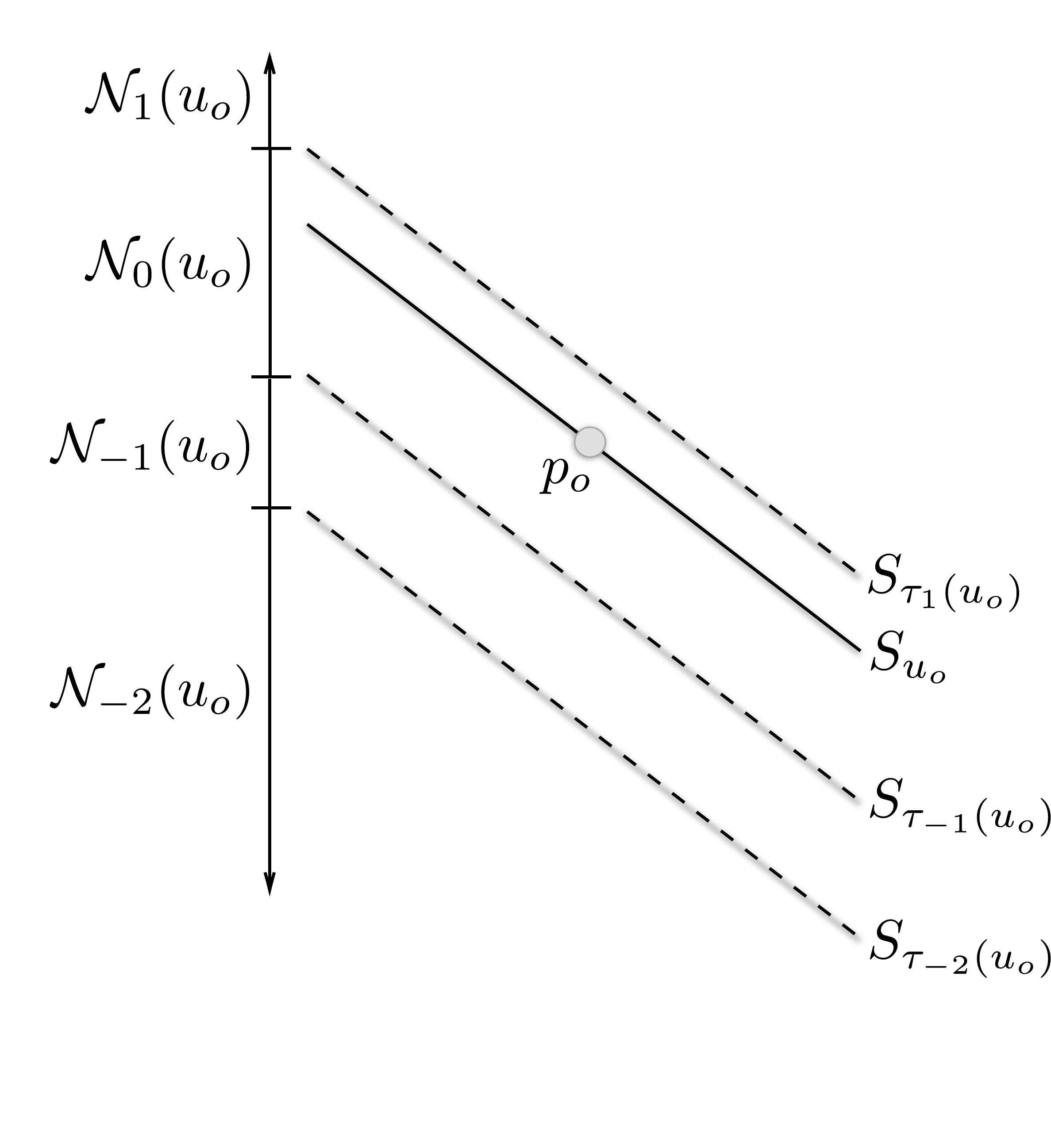}
	
	\vspace{-.8 cm}
	
	\caption{Given a preferred point $p_o$ or a preferred $u = \mbox{constant}$ hyperplane $S_{u_o}$, plane wave spacetimes naturally divide into a number of open four-dimensional regions $\mathcal{N}_n(u_o)$. These regions are separated from each other by the hyperplanes $S_{\tau_n(u_o)}$ conjugate to $S_{u_o}$.}
	
	\label{Fig:Nn}
\end{figure}

In many applications, there exists a preferred point $p_o$, or perhaps a preferred hyperplane $S_{u_o}$. $p_o$ may, for example, represent the position of an observer at a particular time. Fixing this point, the set of hyperplanes conjugate to $S_{u_o}$ divides a plane wave spacetime into a (possibly infinite) number of open regions $\mathcal{N}_n(u_o)$. Let $\mathcal{N}_0(u_o)$ denote the largest connected region containing $S_{u_o}$ and excluding any portion of a hyperplane conjugate to $S_{u_o}$. If there exists a smallest $\tau_1 (u_o) > u_o$ conjugate to $u_o$, the surface $S_{\tau_1(u_o)}$ is clearly contained in the boundary of $\mathcal{N}_0(u_o)$. $\mathcal{N}_1(u_o)$ may then be defined as the largest connected region which includes $S_{\tau_1(u_o)}$ as a boundary and contains points $p_s$ satisfying $u_s > \tau_1(u_o)$ and $I(p_s,p_o) = 0$. This continues the spacetime ``above'' $S_{\tau_1(u_o)}$. Similar constructions may be used to define $\tau_n(u_o)$ and $\mathcal{N}_n(u_o)$ for values of $n$ other than 1. See Fig. \ref{Fig:Nn}. Fixing a particular nonzero integer $n$ and real number $u_o$, it is not necessary that $\tau_n(u_o)$ exist at all. In general, the domain of $\tau_n$ is an open subset of $\mathbb{R}$. This domain can be empty for some $n$.

The geodesic uniqueness result described above can now be reduced to the statement that a point $p_o$ is connected to another point $p_s \neq p_o$ by exactly one geodesic if and only if there exists some $n$ such that $p_s \in \mathcal{N}_n(u_o)$. Two-point tensors like Synge's world function, the parallel propagator, and the van Vleck determinant may be defined unambiguously throughout ``$M \times (\cup_n \mathcal{N}_n)$.'' This excludes from $M \times M$ only a set of measure zero.

\subsection{Geodesics}
\label{Sect:Geodesic}

Beyond the qualitative geodesic structure of plane wave spacetimes discussed above, it is not difficult to obtain explicit coordinate expressions for all geodesics. Let $\Gamma \subset M$ denote some geodesic and $\gamma: \mathbb{R} \rightarrow M$ an affine parametrization of it. The vector field $\ell^a = (\partial/\partial v)^a$ generating the characteristics of the gravitational wave is Killing, so $\dot{\gamma}^a \ell_a$ must be constant on $\Gamma$. If this constant vanishes, $\Gamma$ is confined to a hypersurface of constant phase. Such a geodesic has the form of a (Euclidean) straight line in the coordinates $(v, \bx)$. 

Geodesics satisfying $\dot{\gamma}^a \ell_a  \neq 0$ are more interesting. In these cases, the affine parameter can always be rescaled such that $\dot{\gamma}^a \ell_a = -1$. It is then possible to identify that parameter with the phase coordinate $u$. Doing, so $u ( \gamma(u_s) ) = u_s$ for all $u_s \in \mathbb{R}$. The spatial components $\bm{\gamma} := \bx (\gamma)$ of any geodesic are fixed everywhere once $\bg _o := \bg(u_o)$ and $\dot{\bg}_o := \dot{\bg}(u_o)$ have been specified at some fiducial phase $u_o$. In terms of the matrices $\bA $ and $\bB $ defined in Sect. \ref{Sect:AB},
\begin{equation}
	\bm{\gamma}(u_s) = \bA( u_s,u_o) \bg_o + \bB (u_s,u_o) \dot{\bg}_o. 
	\label{SpatGeo}
\end{equation}
The associated $v$ coordinate of $\Gamma$ may be efficiently derived using that fact that the vector field $(2 v \partial_v + x^i \partial_i)^a$ is a homothety \cite{HarteCaustics}. This implies that
\begin{equation}
	 \fl \qquad \quad v(\gamma(u_s)) = v(\gamma(u_o)) + \frac{1}{2} [\kappa_s (u_s-u_o) + \bg(u_s) \cdot \dot{\bg}(u_s) - \bg_o \cdot \dot{\bg}_o ],
	\label{vGeo}
\end{equation}
where
\begin{equation}
	\kappa_s := - \dot{\gamma}_a \dot{\gamma}^a 
\label{KappaDef}
\end{equation}
is a constant.

$\kappa_s$ is closely related to the conserved quantity on $\Gamma$ associated with the Killing field $\ell^a$. The (unit) 4-velocity $U_s^a$ tangent to $\Gamma$ is related to $\dot{\gamma}^a$ via
\begin{equation}
	U_s^a = \frac{ \dot{\gamma}^a }{ \sqrt{\kappa_s} }.
\label{4Vel}
\end{equation}
It follows that
\begin{equation}
- \ell_a U^a_s = \frac{1}{\sqrt{\kappa_s}}.
\label{lDotU}
\end{equation}
If the spacetime is nearly flat (so $\mathbf{H} \approx 0$), $\kappa_s$ reduces to a particle's specific energy minus its specific momentum in the direction of the gravitational wave.  The limit $-\ell_a U^a \rightarrow 0$ (or $\kappa_s \rightarrow \infty$) may therefore be interpreted as ultrarelativistic motion in the direction of the gravitational wave. By contrast, the limit $-\ell_a U^a \rightarrow \infty$ (or $\kappa_s \rightarrow 0$) corresponds to ultrarelativistic motion \textit{against} the background wave.

Other Killing fields present in essentially all plane wave spacetimes may be written as
\begin{equation}
	(x^i \dot{\Xi}_i ) \ell^a + \Xi^i X^a_{(i)},
\label{Killing}
\end{equation}
where $\Xi_i = \Xi^i$ is any 2-vector with the form
\begin{equation}
\mathbf{\Xi} = \bA (\cdot,u_o) \mathbf{\Xi}(u_o) + \bB (\cdot,u_o) \dot{\mathbf{\Xi}}(u_o).
\end{equation}
For each choice of $u_o$, there exists a four-parameter family of such vector fields. Each of these is associated with a conservation law. Two such conserved quantities may be summarized by
\begin{equation}
	\bm{\mathcal{P}}(u_o) := \frac{1}{ \sqrt{\kappa_s} } \left[ \bA ^\intercal(u_s,u_o) \dot{\bg}(u_s) -  \partial_{(1)} \bA ^\intercal(u_s,u_o) \bm{\gamma}(u_s) \right] .
\label{PDef}
\end{equation}
Fixing any $u_o$, this 2-vector is conserved in the sense that it is independent of $u_s$. Another conserved 2-vector may be defined by
\begin{equation}
	\bm{\mathcal{C}}(u_o) := \frac{1}{ \sqrt{\kappa_s} } \left[  \bB ^\intercal(u_s,u_o) \dot{\bg} (u_s) -  \partial_{(1)} \bB ^\intercal(u_s,u_o) \bm{\gamma}(u_s) \right] .
\label{CDef}
\end{equation}
In the weak-field limit, $\bm{\mathcal{P}}$ corresponds to the specific momentum transverse to the gravitational wave. In this same context, $\bm{\mathcal{C}}$ may be interpreted as the conserved quantity associated with boosts transverse to the gravitational wave. It constrains transverse displacements. Note that both $\bm{\mathcal{P}}$ and $\bm{\mathcal{C}}$ depend on a choice of $u_o$. This is analogous to the choice of origin necessary to define angular momentum in elementary mechanics.

In stationary spacetimes, it is common to discuss various quantities related to gravitational lensing in terms of stationary observers (and often stationary sources). While plane wave spacetimes are not stationary, there does exist sufficient symmetry to define similarly preferred sources and observers. Two geodesics $\Gamma$ and $\Gamma'$ can be said to be ``instantaneously comoving'' at $u=u_o$ when 
\begin{equation}
	\kappa_s = \kappa_s', \qquad \bm{\mathcal{P}}(u_o) = \bm{\mathcal{P}}'(u_o).
	\label{Comoving}
\end{equation}
These conditions imply that the 4-velocities of both geodesics are parallel-transported versions of each other on the constant-phase hyperplane $S_{u_o}$. Note, however, that geodesics which are comoving at one phase are not necessarily comoving at any other phase.

As implied by \eqref{IndexTheorem}, bundles of geodesics are strongly focused on conjugate hyperplanes. Consider such a hyperplane associated with a phase $\tau_n(u_o)$ conjugate to $u_o$ with multiplicity 1. It is clear from \eqref{IDef} that $\hat{\bB}_n(u_o) := \bB (\tau_n(u_o) , u_o) $ is a matrix with rank 1. There therefore exists a unit 2-vector $\hat{\mathbf{q}}_n(u_o)$ such that
\begin{equation}
	\hat{\mathbf{q}}^\intercal_n (u_o) \hat{\bB}_n (u_o) = 0.
	\label{qDef}
\end{equation} 
$\hat{\mathbf{q}}_n(u_o)$ is unique up to sign. Choosing any $\hat{\mathbf{p}}_n(u_o)$ orthogonal to $\hat{\mathbf{q}}_n(u_o)$, the transverse spatial coordinates of all geodesics starting at a given point $p_o$ focus to the line
\begin{equation}
	\bm{\gamma} (\tau_n) = \hat{\bA}_n \bm{\gamma}_o + w \hat{\mathbf{p}}_n
	\label{FocusNonDeg}
\end{equation}
as they pass through $S_{\tau_n(u_0)}$. Here, $w$ is any real number. All $v$ coordinates may be reached on $S_{\tau_n(u_0)}$ by appropriate geodesics. 

Geodesics starting at $p_o$ and intersecting a hyperplane $S_{\tau_n(u_o)}$ with multiplicity 2 all focus to the single transverse position
\begin{equation}
	\bm{\gamma}(\tau_n) = \hat{\bA}_n  \bg_o 
	\label{FocusDeg}
\end{equation}
as they pass through $S_{\tau_n(u_o)}$. As in the multiplicity 1 case, all values of $v$ may be reached by appropriate geodesics. Eqs. \eqref{FocusNonDeg} and \eqref{FocusDeg} illustrate explicitly how conjugate hyperplanes with multiplicities 1 and 2 are associated with astigmatic and anastigmatic focusing, respectively. The former case involves focusing in only one transverse direction, while the latter case involves simultaneous focusing in both directions transverse to the gravitational wave.

\subsection{Distances}
\label{Sect:Synge}

As noted above, all pairs of points not lying on conjugate hyperplanes are connected by exactly one geodesic. There is therefore no ambiguity in ascribing geodesic distances to these pairs. In particular, Synge's world function 
\begin{equation}
	\sigma(p_s,p_o) := \frac{1}{2} (\mbox{squared geodesic distance between $p_s$ and $p_o$}) 
	\label{SigDef0}
\end{equation}
is well-defined whenever its arguments do not lie on conjugate hyperplanes. Plane wave spacetimes constitute one of the few examples where $\sigma$ is known essentially in closed form:
\begin{eqnarray}
	\fl \qquad \sigma (p_s,p_o) &=& \frac{1}{2} (u_s-u_o) \big[ - 2 (v_s-v_o) + \bx^\intercal_s \partial_{(1)} \bB(u_s,u_o) \bB ^{-1}(u_s,u_o) \bx_s
	\nonumber
	\\
	\fl && ~ + 	\bx_o^\intercal \bB ^{-1}(u_s,u_o) \bA(u_s,u_o) \bx_o
	- 2 \bx_o^\intercal \bB ^{-1}(u_s,u_o) \bx_s \big].
	\label{SigDef}
\end{eqnarray}
This is symmetric in its arguments: $\sigma(p_s, p_o) = \sigma(p_o,p_s)$. 

The appearance of $\bB ^{-1}$ in \eqref{SigDef} indicates that $\sigma$ tends to diverge when its arguments approach conjugate hyperplanes. More specifically, suppose that $u_s \approx \tau_n (u_o)$ and $I (\tau_n(u_o), u_o) = 1$. Then,
\begin{equation}
	\sigma(p_s,p_o) \approx - \frac{1}{2} \left( \frac{ u_o - u_s }{ u_s - \tau_n(u_o) } \right) \big[ \hat{\mathbf{q}}_n(u_o) \cdot \big( \bx_s - \hat{\bA}_n(u_o) \bx_o \big) \big]^2
	\label{SigNonDeg}
\end{equation}
is an asymptotic approximation for $\sigma$ if the bracketed term on the right-hand side of this equation is nonzero \cite{HarteCaustics}. It follows from \eqref{FocusNonDeg} that this expression is valid only when there does not exist any geodesic passing from $p_o$ to a point on $S_{\tau_n(u_o)}$ with the same transverse coordinates as $p_s$. The equivalent result if $p_s$ is near a hyperplane $S_{\tau_n(u_o)}$ with multiplicity 2 is
\begin{equation}
	\sigma(p_s,p_o) \approx - \frac{1}{2} \left( \frac{ u_o - u_s }{ u_s - \tau_n(u_o) } \right) \big|  \bx_s - \hat{\bA}_n (u_o) \bx_o \big|^2.
	\label{SigDeg}
\end{equation}
Again, this is valid only when there does not exist any geodesic passing from $p_o$ to a point on $S_{\tau_n(u_o)}$ with the same transverse coordinates as $p_s$.

\section{Image counting}
\label{Sect:NumImages}

One of the most basic questions that can be asked regarding a gravitational lens is the number of images that it produces of a particular source. Stated somewhat differently, how many future-directed null geodesics connect a given timelike curve (the source) to a particular spacetime event (the observer at a particular time)? This may be answered using the geodesic structure of plane wave spacetimes summarized above.

\begin{figure}[t]
	\centering
	
	\includegraphics[width=.3\linewidth]{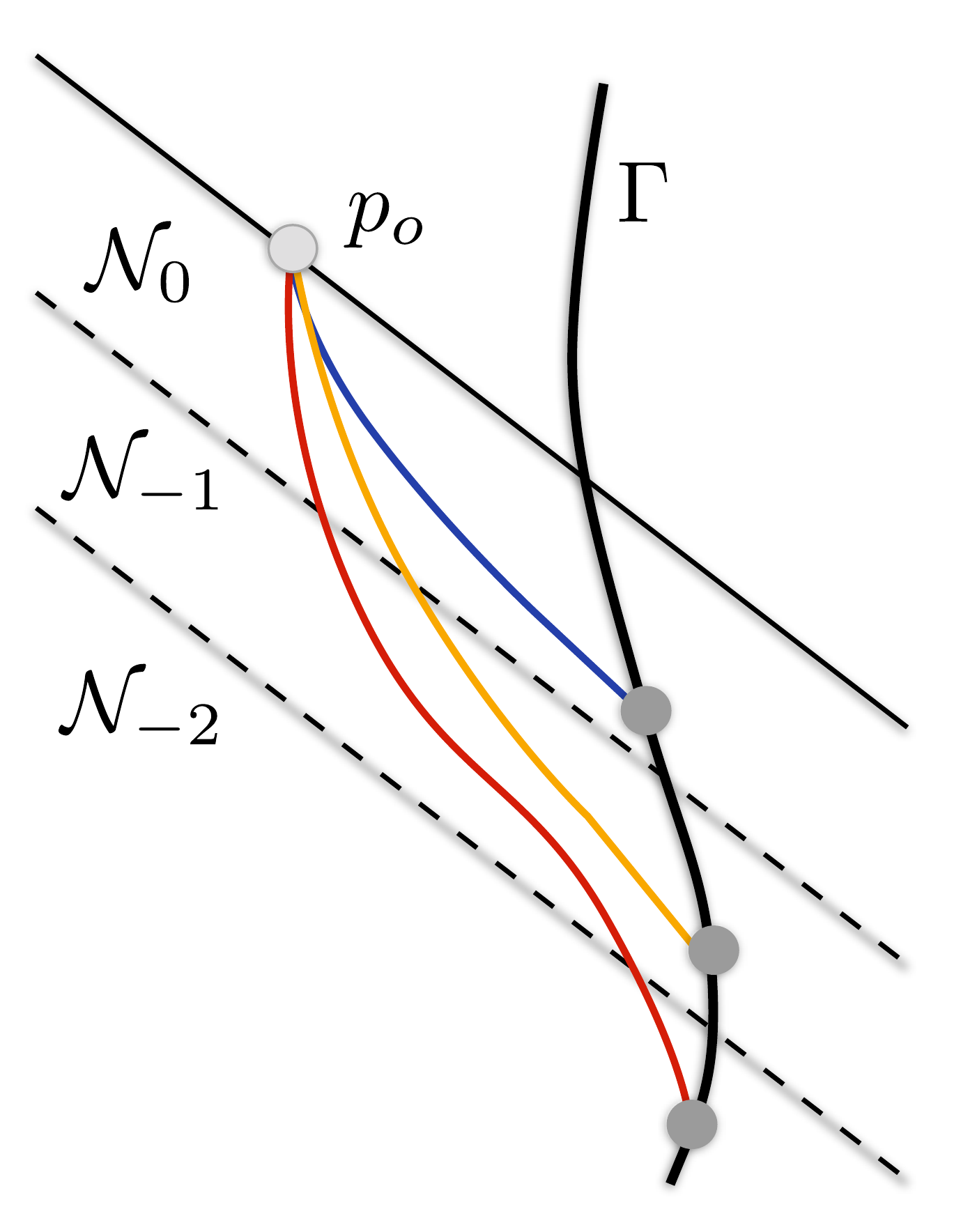}
	
	\caption{Schematic illustration of the gravitational lensing problem. Images correspond to future-directed null geodesics from a timelike source $\Gamma$ to an observation event $p_o$. The arguments of Sect. \ref{Sect:NumImages} show that under generic conditions, exactly one image is emitted from each region $\mathcal{N}_n(u_o)$ lying before the observation event. The dashed lines correspond to constant-$u$ hyperplanes conjugate to $S_{u_o}$. }
	
	\label{Fig:MultImage}
\end{figure}

To fix the notation, let $p_o$ denote a fixed observation event and $\Gamma$ the timelike worldline of a point source. Assume that $\Gamma$ may be parametrized by an everywhere-$C^1$ function $\gamma: \mathbb{R} \rightarrow M$. The phase coordinate $u$ serves as a useful ``quasi-time'' \cite{Ehrlich2} for plane wave spacetimes\footnote{$u$ is not a ``true'' time function because $\nabla_a u$ is null. Generically, there do exist well-behaved functions $t:M \rightarrow \mathbb{R}$ where $\nabla_a t$ is everywhere timelike \cite{HubenyTimeFunction}. These are time functions in the usual sense, although they do not appear to simplify any computations performed here. Incidentally, the existence of such time functions implies that plane wave spacetimes are stably causal. Plane waves are not, however, globally hyperbolic \cite{PenroseHyp}. Hypersurfaces of constant $t$ are not Cauchy surfaces.}, so let $\gamma$ satisfy $u( \gamma(u_s) ) = u_s$ for all $u_s \in \mathbb{R}$. The source's worldline is not required to be geodesic.

This section establishes that under generic conditions, an ideal observer at $p_o$ may see exactly one image of $\Gamma$ from each of the ``epochs'' $\mathcal{N}_n(u_o)$ described in Sect. \ref{Sect:Conjugate}. Somewhat more precisely, there typically exists exactly one future-directed null geodesic from
\begin{equation}
	\Gamma_n := \Gamma \cap \mathcal{N}_n (u_o)
\end{equation}
to $p_o$ for each $n \leq 0$ such that $\mathcal{N}_n(u_o)$ exists. See Fig. \ref{Fig:MultImage}. Recall that the boundaries of $\mathcal{N}_n(u_o)$ depend only on the spacetime under consideration, and not at all on the behavior of any particular source. This allows generic bounds to be placed on time delays associated with the various images that may be observed. 

Recall from Sect. \ref{Sect:Conjugate} that each point in $\Gamma_n$ is connected to $p_o$ via exactly one (not necessarily null) geodesic. Synge's function $\sigma(\gamma(u_s), p_o)$ is therefore well-defined and explicitly given by \eqref{SigDef} for all $u_s$ such that $\gamma(u_s) \in \Gamma_n$.  Consider instead the rescaled function
\begin{equation}
	\Sigma_{n}(u_s) := \frac{\sigma(\gamma(u_s ), p_o)}{u_o-u_s}.
\label{SigCapDef}
\end{equation}
If $n \leq 0$, the domain of $\Sigma_{n}(u_s)$ is equal to all $u_s$ such that $\gamma(u_s) \in \Gamma_n$. If $n=0$, we additionally suppose that $u_s < u_o$ for reasons of causality. Images of $\Gamma_n$ produced by a plane gravitational wave correspond to the zeros of $\Sigma_{n}$. 

Recalling the form \eqref{SigDef} for $\sigma$, it is clear that $\Sigma_{n}$ depends on $\Gamma$ as well as the matrices $\bA $, $\bB $, $\bB ^{-1}$, and $\partial_{(1)} \bB $. We have assumed in Sect. \ref{Sect:AB} that $\bA$ and $\bB $ are at least $C^1$. So is $\gamma$. The definition of $\mathcal{N}_n(u_o)$ ensures that $\det \bB(u_s, u_o) \neq 0$ everywhere $\Sigma_{n}(u_s)$ is defined. $\bB ^{-1}$ is therefore $C^1$ and $\Sigma_{n}$ is continuous. 

$\Sigma_{n}$ is also monotonic. To see this, note that \eqref{EDef}, \eqref{Wronskian}, \eqref{SymMatrices}, \eqref{SigDef}, and \eqref{SigCapDef} may be used to show that
\begin{eqnarray}
\dot{\Sigma}_{n} = \frac{1}{2} \big[ - \dot{\gamma}_a \dot{\gamma}^a +  \big| \dot{\bm{\gamma}} + \bB ^{-\intercal} \left( \bx_o - \partial_{(1)} \bB^\intercal \bm{\gamma} \right) \big|^2 \big] >0.
\label{SigDot}
\end{eqnarray}
$\Sigma_{n}$ is both continuous and monotonic, so at most one zero can exist for each $n$. This means that at most one image of a source may reach an observer from each epoch $\mathcal{N}_n(u_o)$. \textit{Exactly} one such image exists if some $u',u''$ are known to satisfy
\begin{equation}
	\Sigma_n(u')<0, \qquad \Sigma_n(u'')>0.
\end{equation}
Such bounds are easily established.

\subsection{Lensing between conjugate hyperplanes}
\label{Sect:InBetween}

The simplest case to consider is one where $\mathcal{N}_n(u_o)$ lies ``in between'' hyperplanes conjugate to $S_{u_o}$. Suppose that both $\tau_n(u_o)$ and $\tau_{n-1}(u_o)$ exist for some $n<0$. This is true in Fig. \ref{Fig:MultImage} for $n=-1$. More generally, \eqref{SigNonDeg} and \eqref{SigDeg} imply that if there are no geodesics connecting $p_o$ to either $\gamma(\tau_n(u_o))$ or $\gamma(\tau_{n-1}(u_o))$,
\begin{equation}
	\lim_{u_s \rightarrow \tau_{n-1}^+(u_o)} \Sigma_{n} (u_s) = -\infty, \qquad \lim_{u_s \rightarrow \tau_n^-(u_o)} \Sigma_{n} (u_s) = \infty.
\label{SigmaLimits}
\end{equation}
It follows from these limits together with continuity that $\Sigma_{n}$ is surjective on $\mathbb{R}$. Since this function is also monotonic, there must exist exactly one emission phase $u_e \in (\tau_{n-1}(u_o), \tau_n(u_o))$ such that $p_o$ and $\gamma(u_e) \in \Gamma_n$ are connected by a null geodesic. 

Under the same assumptions, projectiles moving on timelike geodesics may be thrown from $\gamma(u_s) \in \Gamma_n$ to $p_o$ only if $u_s < u_e$. Choosing $u_s - \tau_{n-1}(u_o)$ to be sufficiently small (but positive), these projectiles can require an arbitrarily large amount of proper time to intersect $p_o$. It is somewhat curious that points $\gamma(u_s) \in \Gamma_n$ satisfying $u_s > u_e$ cannot be connected to $p_o$ by any causal geodesic. Such points may, however, be reached by suitably accelerated curves which are everywhere causal. 

\subsection{The youngest image}
\label{Sect:Young}

Next, consider the case $n=0$ when there exists at least one conjugate hyperplane in the observer's past (as occurs in the example illustrated by Fig. \ref{Fig:MultImage}). The arguments given above imply that if $\gamma(\tau_{-1}(u_o))$ is geodesically disconnected from $p_o$, 
\begin{equation}
	\lim_{u_s \rightarrow \tau_{-1}^+(u_o)} \Sigma_0 (u_s) = -\infty.
\end{equation}
The other boundary of the domain of $\Sigma_0$ occurs at $u_s = u_o$. Here,
\begin{equation}
	\lim_{u_s \rightarrow u_o^-}  \Sigma_0 (u_s) = \frac{1}{2} \lim_{u_s \rightarrow u_o^-}  \frac{| \bx_o - \bg(u_s) |^2}{u_o-u_s}.
\label{Sigma0}
\end{equation}
This limit clearly tends to $+\infty$ if $\bm{\gamma}(u_o) \neq \bx_o$. Physically, $\bm{\gamma}(u_o) \neq \bx_o$ implies that $p_o$ cannot be connected to $\gamma(u_o)$ by any null geodesic. Assuming that this is true, there must exist exactly one $u_e \in (\tau_{-1}(u_o), u_o)$ such that $p_o$ and $\gamma(u_e) \in \Gamma_0$ are connected by a null geodesic. 

\subsection{The oldest image}
\label{Sect:Oldest}

The results of Sects. \ref{Sect:InBetween} and \ref{Sect:Young} typically suffice to describe the images formed in plane wave spacetimes containing an infinite number of conjugate hyperplanes in an observer's past. It is, however, important to consider cases where only a finite number of conjugate points exist. This occurs, e.g., for finite wavepackets where $\mathbf{H}$ has compact support. It is also true of linearly polarized vacuum waves that are tame in the sense described in \cite{EhrlichTameWave}. 

If there is at least one conjugate hyperplane in the observer's past, let $N$ denote the smallest negative integer such that $\tau_N(u_o)$ exists. If there are no conjugate hyperplanes in the observer's past, set $N=0$. The case illustrated in Fig. \ref{Fig:MultImage} corresponds to $N=-2$ if $\mathcal{N}_{-2}(u_o)$ extends into the infinite past. Regardless, we ask whether there exist any future-directed light rays from $\Gamma_N$ to $p_o$. Unlike in the cases considered above, sources spend an infinite amount of proper time in $\mathcal{N}_N(u_o)$.

Even in flat spacetime, a source that accelerates for an infinitely long time may be causally-disconnected from certain observers via a Rindler horizon. Such phenomena can be ruled out here by supposing that there exists some finite constant $\kappa_{\mathrm{min}} > 0$ such that 
\begin{equation}
	\fl \qquad \qquad - \dot{\gamma}_a \dot{\gamma}^a = 2 \dot{\gamma}^a (u_s) \nabla_a v - \left[ | \dot{\bm{\gamma}}(u_s)|^2 + \bm{\gamma}^\intercal (u_s) \bH(u_s) \bm{\gamma}(u_s) \right] > \kappa_{\mathrm{min}}
\label{vDotBound}
\end{equation}
for all $u_s$ less than some cutoff. As is clear from \eqref{KappaDef} and \eqref{4Vel}, the left-hand side of this inequality acts like the square root of a time dilation factor between the coordinate $u$ and the source's proper time. Eq. \eqref{vDotBound} implies that the source's 4-velocity $U_s^a \propto \dot{\gamma}^a$ satisfies
\begin{equation}
	0 < - \ell_a U^a_s < \frac{1}{\sqrt{\kappa_{\mathrm{min}}}} 
	\label{SourceBound}
\end{equation}
sufficiently far in the past. It therefore excludes sources which experience arbitrarily large boosts against the background gravitational wave in the distant past. It is satisfied by, e.g., sources whose motion is geodesic sufficiently far into the past. Assuming that $\kappa_\mathrm{min}$ exists, it is clear from \eqref{SigDot} that $u'$ may be chosen sufficiently small that
\begin{equation}
	\Sigma_{N}(u') < 0. 
\end{equation}
Further assuming that $\gamma(\tau_N(u_o))$ is geodesically disconnected from $p_o$ (if $N \neq 0$) or that $\gamma(u_o)$ is not null-separated from $p_o$ (if $N=0$), it follows that there exists exactly one $u_e$ smaller than $\tau_N(u_o)$ (if $N\neq 0$) or $u_o$ (if $N=0$) such that $\gamma(u_e) \in \Gamma_N$ is connected to $p_o$ via a future-directed null geodesic.

\subsection{Total image count for generic sources}
\label{Sect:TotImages}

The results just described may be summarized as follows: Suppose that there does not exist any past-directed geodesic segment from $p_o$ to $\Gamma$ whose endpoints are conjugate (in the usual sense). Also assume that the source and observer are not instantaneously aligned with the background wave: $\bx_o \neq \bm{\gamma}(u_o)$. If there exists an ``oldest'' phase conjugate to $u_o$, further require that there be some $\kappa_\mathrm{min} > 0$ such that the source's motion is bounded by \eqref{SourceBound} sufficiently far in the past.

For each $n \leq 0$, these assumptions imply that an observer at $p_o$ sees exactly one image of $\Gamma$ as it appeared in $\mathcal{N}_n(u_o)$. This provides a strong bound on the possible emission times of different images. If $\mathcal{N}_N(u_o)$ exists for every negative integer $N$, an infinite number of images are formed. If, however, there is some smallest $N \leq 0$ such that $\mathcal{N}_N(u_o)$ exists, $|N|+1$ images appear at $p_o$. Note that these results depend only on the waveform $\bH $ and the phase coordinate $u_o$ associated with the observer. The total number of images is the same for all sources satisfying the hypotheses outlined above.

These hypotheses are generic. Recalling \eqref{IndexTheorem}, the requirement that $\Gamma$ exclude any points conjugate to $p_o$ along a connecting geodesic is equivalent to demanding that the source's worldline avoid certain well-behaved one- or two-dimensional subsets of the three-dimensional hyperplanes conjugate to $S_{u_o}$. Similarly, the assumption that $\bx_o \neq \bm{\gamma}(u_o)$ demands only that $\Gamma$ avoid a certain line on $S_{u_o}$. The bound \eqref{SourceBound} on a source's asymptotic 4-velocity can fail to hold only for sources which accelerate for an infinitely long time. Violating any of these conditions for a fixed observer would require that a source's worldline be quite exceptional. Moreover, we now show that $|N|+1$ images appear under even broader (but more difficult to state) conditions than those just discussed.

\subsection{Non-generic imaging}

Despite the comments made above, our assumptions on the behavior of a source's worldline can be violated in certain cases. Suppose, contrary to these assumptions, that there exists at least one point $\gamma_e \in \Gamma$ which is conjugate to $p_o$ along a geodesic connecting these two points. An infinite number of geodesics then pass between $\gamma_e$ and $p_o$. If the connecting geodesics are null, $\gamma_e$ lies on a caustic of the observer's past light cone. Continuous images of point sources -- ``Einstein rings'' -- are then formed at $p_o$ (ignoring the associated breakdown of geometric optics). Such cases are not considered any further here. If conjugate points between the source and observer are associated with non-null geodesics, discrete images of $\Gamma$ appear at $p_o$. In these cases, the methods used above are easily adapted to find how many images of $\Gamma_n$ arrive at $p_o$. As already mentioned, there can be no more than one root for each $\Sigma_n$. Depending on the details of the system, the associated image from $\mathcal{N}_n(u_o)$ may or may not exist.

The case considered in Sect. \ref{Sect:InBetween} where $\Gamma_n$ lies in between successive conjugate hyperplanes is the simplest to analyze. Assuming that $\Gamma$ does not intersect the caustic of $p_o$, no images are formed of $\Gamma_n$ if either $\gamma(\tau_n(u_o))$ is connected to $p_o$ via a timelike geodesic or $\gamma(\tau_{n-1}(u_o))$ is connected to $p_o$ via a spacelike geodesic. Otherwise, exactly one image exists from this region.

If there exists an oldest conjugate hyperplane $S_{\tau_N(u_o)}$ as described in Sect. \ref{Sect:Oldest}, suppose that the source satisfies \eqref{vDotBound} for some $\kappa_\mathrm{min}>0$. There are then zero images of $\Gamma_N$ if $\gamma(\tau_N(u_o))$ is connected to $p_o$ via a timelike geodesic. There is exactly one image if these points are either geodesically disconnected or are connected by a spacelike geodesic.

The last cases to consider concern images of $\Gamma_0$. First suppose that $\bx_o \neq \bm{\gamma}(u_o)$. If there exists at least one conjugate hyperplane in the observer's past, one image is formed of $\Gamma_0$ if either $\gamma(\tau_{-1}(u_o))$ is geodesically disconnected from $p_o$ or it is connected by a timelike geodesic. No images are formed if  $\gamma(\tau_{-1}(u_o))$ and $p_o$ are connected by a spacelike geodesic. If $\bx_o \neq \bm{\gamma}(u_o)$ and there are no conjugate points in the observer's past, condition \eqref{vDotBound} implies that there exists exactly one image of $\Gamma_0$.

Cases where the source and observer are instantaneously aligned are more interesting. Suppose that $\bx_o = \bm{\gamma}(u_o)$. There then exists one image of $\Gamma_0$ with $u_e = u_o$. Recalling that $u_o$ is not in the domain of $\Sigma_0$, it is possible for a second image to be emitted from $\Gamma_0$ if $\Sigma_0 = 0$ somewhere. This may be seen by noting that
\begin{equation}
	\lim_{u_s \rightarrow u_o^-} \Sigma_{0} = v ( \gamma(u_o)) - v_o.
\end{equation}
Two images of $\Gamma_0$ can therefore exist when $v(\gamma(u_o)) > v_o$ and $\bx_o = \bm{\gamma}(u_o)$. 

If a source includes points which are conjugate to the observer (in the ordinary sense), there is no simple result for the total number of images formed. Nevertheless, it is always possible to say that the total number of images is less than or equal to $|N|+2$ if a source does not intersect a caustic of the observer's light cone.

\section{Properties of lensed images}
\label{Sect:ImageProperties}

Plane wave spacetimes typically produce multiple images of each source. Even for sources whose intrinsic properties remain constant, these images can appear with different spectra, brightnesses, etc. We now compute these properties for generic configurations satisfying the hypotheses summarized in Sect. \ref{Sect:TotImages}. 

For each image of a timelike worldline $\Gamma$ seen at $p_o$, there is an associated null geodesic segment connecting $p_o$ to an appropriate emission point $\gamma_e = \gamma(u_e) \in \Gamma$. In terms of Synge's function \eqref{SigDef0}, these points satisfy
\begin{equation}
	\sigma (\gamma_e, p_o) = 0.
\label{SigZero}
\end{equation}
First derivatives of $\sigma$ are always tangent to the connecting light ray. In particular, the vector
\begin{equation}
	r^a_o  := -\frac{ \nabla^{a} \sigma ( \gamma_e, p_o) }{ u_o - u_e}
\label{kDef}
\end{equation}
at $p_o$ points along the geodesic which eventually intersects $\gamma_e$ (and is therefore past-directed). The derivative operator here is understood to act on the second argument of $\sigma$. Also note that $r^a_o$ is normalized such that $\ell_a r^a_o = 1$. Parallel-transporting $r^a_o$ to the observation point yields
\begin{equation}
	r^a_e  =    \frac{ \nabla^a \sigma ( \gamma_e, p_o) }{ u_o - u_e}.
\label{kDef2}
\end{equation}
The derivative operator in this equation is understood to act on the first argument of $\sigma$. Both $r^a_o$ and $r^a_e$ may be viewed as (dimensionless) separation vectors between $p_o$ and $\gamma_e$. 

Eq. \eqref{SigDef} and the various identities of Sect. \ref{Sect:AB} may be used to compute the explicit coordinate components of $r^a_e$ and $r^a_o$. Components transverse to the direction of wave propagation are
\begin{equation}
	\mathbf{r}_e =  \bB ^{-\intercal}(u_e,u_o) \left[  \bx_o - \partial_{(1)} \bB^\intercal (u_e, u_o) \bg_e   \right],
\label{Tangent1}
\end{equation}
where $\bx_o = \bx(p_o)$ and $\bg _e = \bx(\gamma(u_e))$ denote the transverse coordinates of the observer and source. A similar calculation shows that
\begin{equation}
	\mathbf{r}_o =  \bB ^{-1}(u_e,u_o) \left[ \bA (u_e, u_o) \bx_o - \bg_e \right].
\label{Tangent2}
\end{equation}
$\mathbf{r}_o$ and $\mathbf{r}_e$ are related via
\begin{equation}
	\mathbf{r}_e = \partial_{(1)} \bB (u_e, u_o) \mathbf{r}_o - \partial_{(1)} \bA (u_e, u_o) \bx_o.
\label{TangentDiff}
\end{equation}

Much of the discussion below considers sources moving on geodesics. In these cases, use of \eqref{SpatGeo} shows that
\begin{equation}
	\bR_o =  \bB ^{-1}(u_e,u_o) \bA (u_e,u_o) \delta \bx_o - \dot{\bg}_o.
\label{LambdaDotGeo}
\end{equation}
Here, $\delta \bx_o := \bx_o - \bg (u_o) = \bx_o - \bg _o$. The various identities involving $\bA $ and $\bB $ discussed in Sect. \ref{Sect:AB} may also be used to reduce the imaging condition \eqref{SigZero} to
\begin{eqnarray}
	\fl \qquad \qquad \kappa_s (u_o - u_e ) &=& 2(\dot{\bg}_o  \cdot \delta \bx_o - \delta v_o) -  \delta \bx_o^\intercal \bB ^{-1} (u_e,u_o) \bA (u_e, u_o) \delta \bx_o .
\label{SigGeo}
\end{eqnarray}
Here, $\delta v_o := v(p_o) - v(\gamma_o)$. Eq. \eqref{SigGeo} is a nonlinear relation for the emission ``time'' $u_e$ in terms of the observer's position $p_o$ and the parameters $\bg_o$, $\dot{\bg}_o$, $v(\gamma_o)$, $\kappa_s$ describing the source's worldline. As discussed in Sect. \ref{Sect:NumImages}, there can be many solutions to \eqref{SigGeo}. These correspond to different images.

Neither $u_e$ nor $r^a_o$ depends on the observer's motion. Nevertheless, redshifts and angles on the observer's sky do depend on that motion (as is true even in flat spacetime). It is often useful to fix this effect by supposing that the observer is instantaneously comoving with the source. Following  \eqref{Comoving}, this is taken to mean that the unit 4-velocities $U^a_s, U^a_o$ of the source and observer on $S_{u_o}$ satisfy
\begin{eqnarray}
	\ell_a U_s^a(u_o)  = \ell_a U^a_o  = - \frac{1}{ \sqrt{\kappa_s} } , 
	\label{Comoving2}
	\\
	\bm{\mathcal{P}}_s (u_o) = \mathbf{U}_s (u_o) = \mathbf{U}_o = \frac{ \dot{\bg}(u_o) }{ \sqrt{\kappa_s} } = \frac{ \dot{\bx}_o }{ \sqrt{\kappa_s} }.
	\label{Comoving3}
\end{eqnarray} 
Recall from Sect. \ref{Sect:Geodesic} that the ``transverse momentum'' $\bm{\mathcal{P}}_s(u_o)$ is generated by contracting $U^a_s(u_o)$ with the two Killing fields equal to $X^a_{(i)}$ at $p_o$ and having vanishing first derivative at that point. Also note that \eqref{Comoving3} implicitly defines an instantaneous observer velocity $\dot{x}^a_o = \sqrt{\kappa_s} U^a_o$ normalized (like $\dot{\gamma}^a$) such that $\ell_a \dot{x}^a_o = -1$.

\subsection{Frequency shifts}
\label{Sect:Redshift}

Gravitational lenses typically discussed in astrophysics involve nearly-Newtonian mass distributions which may be regarded as approximately stationary (at least on sub-cosmological timescales). If both a source and an observer are sufficiently far from such a lens, there can be no significant redshift or blueshift from the gravitational field of that lens. Roughly speaking, a light ray falling into any stationary gravitational potential must climb out of that same potential. This result breaks down if light passes through non-stationary regions of spacetime. Indeed, plane wave spacetimes may produce images with significant frequency shifts \cite{FaraoniRedshift}.

Consider an approximately monochromatic beam of light emitted from $\gamma_e$ and received at $p_o$. A future-directed tangent vector $k^a_e \propto -r^a_e$ to the emitted light ray may always be chosen such that
\begin{equation}
	\omega_e = - k_e \cdot U_e
\end{equation}
is the angular frequency of the light as seen by its source. The frequency $\omega_o$ of this same light ray as measured by an observer at $p_o$ is $-k_o \cdot U_o$, where $k_o^a$ is equal to $k^a_e$ parallel transported from the source to the observer. The observed and emitted frequencies are therefore related by
\begin{equation}
\frac{\omega_o}{\omega_e}  = \frac{ U_o \cdot k_o}{  U_e \cdot k_e  }=  \frac{ U_o \cdot r_o}{  U_e \cdot r_e  } = \sqrt{ \frac{ \kappa_e }{ \kappa_o } }  \left( \frac{ \kappa_o + | \dot{\bx}_o + \bR _o |^2 }{ \kappa_e + | \dot{\bg}_e + \bR _e |^2  } \right).
\label{RedshiftGen}
\end{equation}
Here, $\kappa_e :=1/(\ell \cdot U_s(u_e))^2$ and $\kappa_o :=1/( \ell \cdot U_o)^2$. The 2-vectors $\bR _e$ and $\bR _o$ appearing here are determined by the source and observer positions via \eqref{SigDef}, \eqref{SigZero}, \eqref{Tangent1}, and \eqref{Tangent2}. The resulting frequency shift is valid for all emission points not contained in a caustic of $p_o$.

Now suppose that a source moves on a geodesic and that the observer is instantaneously comoving with this geodesic in the sense of \eqref{Comoving2} and \eqref{Comoving3}. Then $\kappa_e = \kappa_s = 1/(\ell \cdot U_s)^2$ doesn't depend on which image is chosen. Eqs. \eqref{TangentDiff}, \eqref{LambdaDotGeo}, and the symmetry of $\bB ^{-1} \bA $ may be used to rewrite \eqref{RedshiftGen} as
\begin{equation}
	\frac{ \omega_o }{ \omega_e } = 1 + (\bB ^{-\intercal} \delta \bx_o )^\intercal  \left( \frac{  \bA\bA ^\intercal - \bm{\delta} }{ \kappa_s + | \bB ^{-\intercal} \delta \bx_o |^2} \right) (\bB ^{-\intercal} \delta \bx_o).
\label{RedshiftGeodesic}
\end{equation}
The matrix in parentheses on the right hand side of this equation acts like a metric for the ``separation'' 2-vector $\bB ^{-\intercal}(u_e,u_o) [\bx_o - \bg (u_o)]$. If both eigenvalues of $\bA(u_e, u_s) \bA ^\intercal (u_e,u_s) - \bm{\delta}$ are negative, the source is necessarily redshifted. Conversely, sources are always blueshifted when this matrix is positive definite. If $\bA \bA ^\intercal - \bm{\delta}$ has both positive and negative eigenvalues, the sign of the frequency difference depends on the direction of $ \bB ^{-\intercal} \delta \bx_o$. For special configurations, there is no frequency shift at all.

\subsection{Angles}
\label{Sect:Angles}

Various images formed from a single source appear at different points on an observer's sky. Like redshifts, the relative angles between images change depending on an observer's 4-velocity $U^a_o$. The angle $\theta$ between two images arriving at $p_o$ with tangents $r^a_o$ and $r'^a_o$ is
\begin{equation}
\cos \theta = \frac{ (g_{ab} + U_{o,a} U_{o,b}) r^a_o r'^b_o}{ (U_o \cdot r_o) (U_o \cdot r_o') } = 1 + \frac{ r_o \cdot r_o' }{ (U_o \cdot r_o) (U_o \cdot r_o')} .
\end{equation}
Simplifying,
\begin{equation}
\cos \theta = 1 - \frac{ 2 \kappa_o | \bR _o - \bR _o' |^2 }{\left( \kappa_o + | \dot{\bx}_o + \bR_o  |^2 \right)  \left( \kappa_o+ | \dot{\bx}_o + \bR_o'  |^2 \right)  }.
\label{AnglesGen}
\end{equation}
This expression is valid for arbitrary source and observer configurations. Specializing to geodesic sources and comoving observers,
\begin{eqnarray}
	\cos \theta &=& 1 - \frac{ 2 \kappa_s \left| (\bB ^{-1} \bA - \bB '^{-1} \bA') \delta \bx_o \right|^2 }{ (\kappa_s + | \bB ^{-1} \bA \delta \bx_o|^2) (\kappa_s + | \bB '^{-1} \bA' \delta \bx_o|^2 ) } .
	\label{CosTheta}
\end{eqnarray}
Here, $\bA = \bA (u_e,u_o)$ and $\bA ' = \bA (u_e',u_o)$. It is evident that angles are largely controlled by the difference between $\bB ^{-1} \bA$ at the two emission times. 

Another interesting angle to consider is the observed separation $\psi$ between a single image (emitted at $\gamma_e$) and a generator $\ell^a$ of the background gravitational wave. For arbitrarily moving source and observer configurations,
\begin{equation}
	\cos \psi =  1- \frac{ 2 \kappa_o }{ \kappa_o + | \dot{\bx}_o + \bR _o|^2} .
\end{equation}
For observers comoving with geodesic sources,
\begin{equation}
	\cos \psi = 1 - \frac{2 \kappa_s}{\kappa_s + | \bB ^{-1} \bA \delta \bx_o |^2  }.
	\label{PsiDef}
\end{equation}
This may be used to rewrite the angle $\theta$ between two different images partially in terms of the angles $\psi$ and $\psi'$ those images make with $\ell^a$. Using \eqref{CosTheta}, 
\begin{equation}
	\cos \theta = \cos \psi \cos \psi' + \frac{ (\bB ^{-1} \bA \delta \bx_o) \cdot (\bB '^{-1} \bA' \delta \bx_o) }{ |\bB ^{-1} \bA \delta \bx_o| |\bB '^{-1} \bA' \delta \bx_o| } \sin \psi \sin \psi'  .
\end{equation}
Similarly, the frequency shift \eqref{RedshiftGeodesic} of an individual image may be rewritten as
\begin{equation}
	\frac{ \omega_o }{ \omega_e }= \frac{ \kappa_s \csc^2 (\psi/2) }{ \kappa_s + | \bB ^{-\intercal} \delta \bx_o|^2}.
	\label{AngleRedshift}
\end{equation}
It is evident from this equation that images which appear highly blueshifted to comoving observers must satisfy $\psi \approx 0$.

\subsection{Image distortion and magnification}
\label{Sect:ImageDist}

\begin{figure}
	\centering
	\includegraphics[width=.4\linewidth]{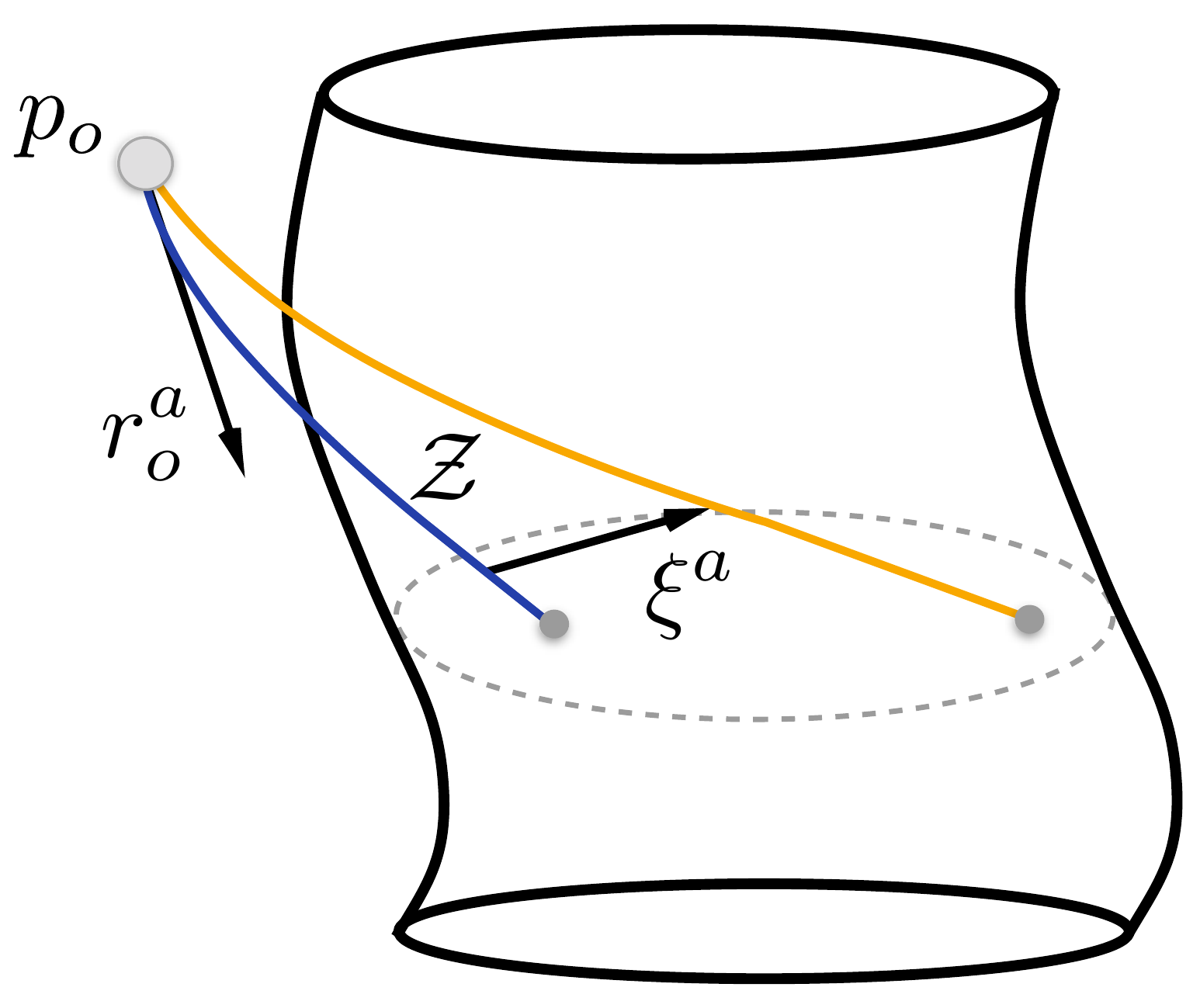}
	
	\caption{Imaging for an extended source. The fiducial light ray $\mathcal{Z}$ is indicated together with another light ray separated from it by a deviation vector $\xi^a$. The vector $r^a_o$ is also drawn. This is tangent to $\mathcal{Z}$ at the observation point $p_o$.}
	
	\label{Fig:Distortion}
\end{figure}

Thus far, all sources here have been modelled as though they were confined to timelike worldlines. Real objects are not pointlike, however. They form extended worldtubes in spacetime. Images of such worldtubes form null geodesic congruences which converge on $p_o$. These images can be significantly distorted by the curvature of spacetime. It is simplest to quantify such distortions by first fixing a particular null geodesic $\mathcal{Z}$ passing between some part of the source and $p_o$. Precisely which geodesic is chosen is not important. $\mathcal{Z}$ serves only as an origin from which to discuss nearby light rays connecting $p_o$ to other points in the source. Once this origin has been fixed, the image of an extended source may be described entirely using deviation vectors on $\mathcal{Z}$ (at least for sufficiently small sources). See Fig. \ref{Fig:Distortion}.

Deviation vectors (or Jacobi fields) satisfy the geodesic deviation (or Jacobi) equation along $\mathcal{Z}$. Letting $r^a$ denote the past-directed null vector tangent to $\mathcal{Z}$ and obtained by parallel-transporting $r^a_o$ from $p_o$, every deviation vector $\xi^a$ is a solution to
\begin{equation}
	r^b \nabla_b (r^c \nabla_c \xi^a) = R^{a}{}_{bcd} r^b r^c \xi^d. 
\end{equation}
This equation is linear, so $\xi^a$ must depend linearly on initial data. In particular, all deviation vectors can be written as linear combinations of appropriate bitensors contracted into the initial data\footnote{Capital letters are used in this subsection to denote abstract indices associated with the observation point $p_o$. This is done to avoid confusion when writing down two-point tensors such as $B^{a}{}_{A}$ [see \eqref{BDef}].} $\xi^A(u_o)$ and $\dot{\xi}^A(u_o)$. All light rays observed at $p_o$ must necessarily intersect that point, so it suffices to set $\xi^A(u_o) = 0$. The first derivative of a deviation vector at $p_o$ describes an angular deviation between one point of an image and the center associated with $\mathcal{Z}$. We therefore consider deviation vectors $\xi^a$ with the form
\begin{equation}
	\xi^a = B^{a}{}_{A} (\cdot, u_o) \dot{\xi}^{A}(u_o).
	\label{BDef}
\end{equation}
$B^{a}{}_{A}$ is known as a Jacobi propagator. It satisfies the Jacobi equation
\begin{equation}
	r^{b}\nabla_{b} ( r^{c} \nabla_{c} B^{a}{}_{A} ) = R^{a}{}_{bcd} r^{b} r^{c} B^{d}{}_{A}
		\label{Jacobi}
\end{equation}
along $\mathcal{Z}$ together with the initial conditions
\begin{equation}
 	\lim_{u_s \rightarrow u_o} B^{a}{}_{A}(u_s,u_o)  = 0, \quad \lim_{u_s \rightarrow u_o} r^{b} \nabla_{b} B^{a}{}_{A}(u_s,u_o) =\delta^{a}_{A}.
 \end{equation}
Note that $B^{a}{}_{A}$ is a bitensor. It maps vectors at $p_o$ into vectors at others points on $\mathcal{Z}$. The transverse components of $B^{a}{}_{A}$ are
\begin{equation}
	B_{aA} X^a_{(i)} X^A_{(j)} = (\bB )_{ij},
	\label{BTrans}
\end{equation}
where $\bB $ is the matrix defined in Sect. \ref{Sect:AB}. Other components of $B^{a}{}_{A}$ may be deduced from the eigenvector relations \cite{HarteCaustics}
\begin{eqnarray}
	\fl \qquad \quad B^{a}{}_{A}(u_s,u_o) r_o^A = (u_s - u_o) r^a, \qquad r^a B_{a}{}^{A} (u_s,u_o)= (u_s-u_o) r_{o}^A,
\label{rDotB}
\\
	\fl \qquad \quad B^{a}{}_{A}(u_s,u_o)  \ell^A = (u_s - u_o) \ell^a ,  \qquad \ell_a B^{a}{}_{A}(u_s,u_o) = (u_s-u_o) \ell_A.
	\label{ellDotB}
\end{eqnarray}

All parts of an image must arrive at an observer along null geodesics. Additionally, an observer with 4-velocity $U^a_o$ can only measure angles of vectors orthogonal to $U^a_o$. It therefore suffices to restrict attention to deviation vectors satisfying
\begin{equation}
	r_o \cdot \dot{\xi}(u_o) = U_o \cdot \dot{\xi}(u_o)=0
\end{equation}
at $p_o$. These constraints restrict all interesting initial data to a two dimensional space. The orthonormal vectors
\begin{equation}
	e_{(i)}^A = X^A_{(i)} - r_o^i \ell^A - 2 \left( \frac{  r_o^i + \dot{x}_o^i }{ \kappa_o + | \dot{\bx}_o + \bR _o|^2} \right) r_o^A
	\label{SachsO}
\end{equation}
form a basis for this space at $p_o$. They satisfy
\begin{equation}
	e_{(i)} \cdot e_{(j)} = \delta_{ij}, \qquad r_o \cdot e_{(i)} = U_o \cdot e_{(i)} = 0.
\end{equation}
Parallel-transporting $e^A_{(i)}$ to another point on $\mathcal{Z}$ yields
\begin{equation}
	e_{(i)}^a = X^a_{(i)} - r^i \ell^a - 2 \left( \frac{  r_o^i + \dot{x}_o^i }{ \kappa_o + | \dot{\bx}_o+ \bR _o|^2} \right) r^a,
	\label{SachsGen}
\end{equation}
which forms a Sachs basis \cite{PerlickReview} on $\mathcal{Z}$. Initial data appearing in \eqref{BDef} must be of the form
\begin{equation}
	\dot{\xi}^A (u_o) = (U_o \cdot r_o) \alpha^{(i)} e^A_{(i)} ,
	\label{InitData}
\end{equation}
where $\bm{\alpha}$ is an unconstrained 2-vector. The factor $(U_o \cdot r_o) >0$ is included here so that $\bm{\alpha}$ is directly related to angles on an observer's sky. A sufficiently small image may be described by a suitable set of 2-vectors $\bm{\alpha}$ representing the angular locations of each portion of the image with respect to the center defined by $\mathcal{Z}$. 

Each $\bm{\alpha}$ may be translated into a physical displacement at the source using \eqref{BDef} and \eqref{InitData}. First note that for every particular $\bm{\alpha}$, \eqref{rDotB} implies that $r \cdot \xi = 0$ throughout $\mathcal{Z}$. Indeed, $\xi^a$ is always a linear combination of the $e^a_{(i)}$ together with $r^a$. Components of $\xi^a$ proportional to $r^a$ are physically irrelevant, so we consider only the Sachs components
\begin{equation}
	\xi_{(i)} := e_{(i)} \cdot \xi = [(U_o \cdot r_o)  B_{aA} e^a_{(i)} e^A_{(j)} ] \alpha^{(j)}.
\end{equation}
Defining the $2 \times 2$ matrix
\begin{equation}
	D_{(i)(j)}(u_s,u_o) :=  (U_o \cdot r_o) B_{aA}(u_s,u_o) e^a_{(i)}(u_s) e^A_{(j)}(u_o),
	\label{DDef}
\end{equation}
it is then clear that $\bm{\xi} = \mathbf{D} \bm{\alpha}$ for any $\bm{\alpha}$. $\mathbf{D}$ is referred to as the Jacobi matrix or Jacobi map \cite{PerlickReview,EhlersShape}. Using \eqref{BTrans}-\eqref{ellDotB},  \eqref{SachsO}, \eqref{SachsGen}, and \eqref{DDef},
\begin{equation}
	\mathbf{D} (u_s,u_o) = (U_o \cdot r_o) \bB (u_s , u_o).
\end{equation}
If a source moves on a geodesic which is instantaneously comoving with the observer, 
\begin{equation}
	U_o \cdot r_o = \frac{ \kappa_s+|\bB ^{-1} \bA \delta \bx_o|^2}{ 2 \sqrt{\kappa_s} }. 
\end{equation}

This discussion implies that a portion of an image with angular separation $\epsilon \bm{\alpha}$ from the fiducial direction associated with $r_o^a$ is spatially separated from the fiducial emission point $\gamma_e \in \mathcal{Z}$ by
\begin{equation}
	\epsilon \bm{\xi}(u_e) = (U_o \cdot r_o) \bB (u_e,u_o) (\epsilon \bm{\alpha}).
\label{BInterpret}
\end{equation}
The factors of $\epsilon \ll 1$ have been introduced here to emphasize that this description is valid only for infinitesimal deviations. Regardless, \eqref{BInterpret} shows that up to the time dilation factor $(U_o \cdot r_o)$, the matrix $\bB $ central to all aspects of plane wave geometry may be physically interpreted as a transformation converting infinitesimal angles on the vertex of a light cone into infinitesimal separations elsewhere on that light cone. $\bB $ depends only on the $u$ coordinates of the source and emission points, and not on any other aspects of the physical configuration. It may be computed for all possible observer-source pairs directly from the wave profile $\bH $. 

Angles of emission (as opposed to observation) of the various light rays travelling from the source to the observer may be found by differentiating \eqref{BInterpret} and applying the appropriate time dilation factor:
\begin{equation}
	\frac{\dot{\bm{\xi}} (u_e)}{ U_e \cdot r_e }  = \left( \frac{ U_o \cdot r_o }{ U_e \cdot r_e } \right) \partial_{(1)} \bB(u_e,u_o) \bm{\alpha} = \left( \frac{ \omega_o} {\omega_e} \right) \partial_{(1)} \bB(u_e,u_o) \bm{\alpha}.	
	\label{AngleEmission}
\end{equation}
The last equality here makes use of \eqref{RedshiftGen}. Applying \eqref{BInterpret} shows that
\begin{equation}
	\dot{\bm{\xi}} = \partial_{(1)} \bB \bB ^{-1} \bm{\xi}.
\end{equation}
The symmetric matrix $\partial_{(1)} \bB \bB ^{-1}/(U_e \cdot r_e)$ therefore converts spatial locations to emission angles within the source (with the constraint that all light rays intersect $p_o$). It is referred to as the optical deformation matrix \cite{PerlickReview,EhlersShape}. 

Eq. \eqref{BInterpret} implies that there is a sense in which circles on the observer's sky correspond to ellipses near $\gamma_e$. This deformation may be parametrized by performing a polar decomposition on $\mathbf{D}(u_e,u_o)$:
\begin{equation}
	 \mathbf{D} = \mathbf{R}_\beta^\intercal
			\left( \begin{array}{cc}
				D_{+}	&	0	\\
				0	&	D_{-}
			\end{array} \right) \mathbf{R}_\chi.
	\label{Distortion}
\end{equation}
Here, $\mathbf{R}_\beta$ and $\mathbf{R}_\chi$ represent rotation matrices through some angles $\beta$ and $\chi$. The ratio $D_+/D_-$ is related to the ellipticity of the aforementioned ellipse. $\chi$ represents the angle between the principal axes of that ellipse and the Sachs basis. $D_\pm$ and $\chi$ are referred as shape parameters \cite{PerlickReview,NewmanImageDist}.

Recalling that polarization vectors are parallel-transported in the geometric optics approximation \cite{EhlersBook}, any polarization vector must have Sachs components which are constant along $\mathcal{Z}$. In principle, the angle $\chi$ might therefore be measured by comparing the relative ``rotation'' between an object's observed shape and an appropriate polarization angle \cite{PerlickReview, Polarization}. For linearly polarized waves where $\bH $ can be made diagonal by an appropriate coordinate choice, $\chi =0$ with respect to this coordinate system and the basis \eqref{SachsGen}. It is shown in Sect. \ref{Sect:WeakWave} that $\chi$ also vanishes in a natural way for all sufficiently weak wavepackets which are nonzero only for short times.

Eq. \eqref{BInterpret} implies that $\mathbf{D} $ converts angles at the observer to separations within the source.  The determinant of $\mathbf{D}$ must therefore relate solid angles at $p_o$ to physical areas near $\gamma_e$:
\begin{equation}
	\frac{ \rmd \mathcal{A} }{ \rmd \Omega } = |\det \mathbf{D} |  = | D_+ D_- |  = (U_o \cdot r_o)^2 | \det \bB  |.
\end{equation}
It follows that
\begin{equation}
	d_\mathrm{ang} := \sqrt{ | \det \mathbf{D} | } = (U_o \cdot r_o) \sqrt{ | \det \bB | }
	\label{Distance}
\end{equation}
may be interpreted as an ``angular diameter distance.'' Absolute value signs are necessary here because $\det \bB $ changes sign after each pass through a conjugate hyperplane with multiplicity 1. Physically, such sign changes represent parity inversions of the resulting image. Note that $d_\mathrm{ang}$ does not necessarily increase monotonically with the age of an image (as computed using the source's proper time). 

Closely related to the angular diameter distance is the luminosity distance
\begin{equation}
	d_\mathrm{lum} := (\omega_o/\omega_e)^{-2} d_\mathrm{ang}.
	\label{dLum}
\end{equation}
One factor of $\omega_o/\omega_e$ arises here from considering light cones emanating from the source instead of the observer. The other factor of $\omega/\omega_e$ is related to the energy change associated with frequency shifts.

\section{Symmetric plane waves}
\label{Sect:SymWaves}

Now that various optical quantities have been computed for general plane wave spacetimes, we consider their application to various special cases. The simplest nontrivial plane waves are the symmetric waves. These are locally symmetric in the sense that $\nabla_a R_{bcd}{}^f=0$. It follows from \eqref{Riemann} that symmetric plane waves must have constant waveforms. Also note that $(\partial/\partial u)^a$ is Killing in these examples [as well as $\ell^a = (\partial/\partial v)^a$, which is Killing in all plane wave spacetimes]. Particular symmetric plane waves may be specified entirely by the (constant) eigenvalues of $\bH $. 

Recalling the decomposition \eqref{HGen} of $\bH $ into $h_+$, $h_\times$ and $h_\|$, a coordinate rotation may always be used to set $h_\times =0$ for symmetric waves. It is then evident that the two eigenvalues of $\bH $ are given by $\pm h_+ - h_\| $. It is always possible to set
\begin{equation}
\bH = \left( \begin{array}{cc}
-h_1	&	0	\\
0	&	-h_2
\end{array} \right),
\label{SymWave}
\end{equation}
where
\begin{equation}
	h_{1} := h_\| + h_+, \qquad h_2 := h_\| - h_+ .
\end{equation}
The weak energy condition implies that $h_\| \geq 0$, so at least one eigenvalue of $\bH $ must be negative (implying that at least one of the $h_{1,2}$ must be positive). We assume for definiteness that $h_+ \geq 0$. Then,
\begin{equation}
	h_1 > 0, \qquad h_1 \geq |h_2|.
\end{equation}
If the vacuum Einstein equation is imposed, $h_\| =0$ and $h_2 = -h_1$. For conformally-flat geometries representing spacetimes associated with, e.g., pure electromagnetic plane waves, $h_+ = 0$ and $h_2 = h_1$. Other cases may be viewed as superpositions of gravitational and (``gravito''-)electromagnetic waves. 

All symmetric waves produce an infinite number of images of almost every source. It is clear from \eqref{SymWave} that these waves are also linearly polarized. The angles $\chi$ and $\beta$ appearing in \eqref{Distortion} therefore vanish when considering image deformations with respect to the Sachs basis \eqref{SachsGen}. Other lensing properties depend on the sign of $h_2$. We call the case $h_2< 0$ ``gravity-dominated'' and the case $h_2>0$ ``matter-dominated.'' 

\subsection{Gravity-dominated symmetric waves}

Consider symmetric plane wave spacetimes where $h_2 = h_\| - h_+ < 0$. Gravity-dominated waves such as these generalize the vacuum waves satisfying $h_1 = - h_2$. Symmetric vacuum waves arise from, e.g., the Penrose limit of a null geodesic orbiting a Schwarzschild black hole on the light ring. 

For any gravity-dominated symmetric wave, the matrices $\bA $ and $\bB$ defined in Sect. \ref{Sect:AB} are
\begin{eqnarray}
\fl \qquad \quad \bA (u_s,u_o) &=& \left( \begin{array}{cc}
\cos  h_1^{ \frac{1}{2} } (u_s-u_o)	&	0	\\
0	&	\cosh |h_2|^{ \frac{1}{2} } (u_s-u_o)
\end{array} \right),
\\
\fl \qquad \quad \bB (u_s,u_o) &=& \left( \begin{array}{cc}
h_1^{ - \frac{1}{2} } \sin h_1^{ \frac{1}{2} } (u_s-u_o)	&	0	\\
0	&	|h_2|^{ -\frac{1}{2} }  \sinh |h_2|^{ \frac{1}{2} } (u_s-u_o)
\end{array} \right).
\end{eqnarray}
It is clear that $\det \bB (\cdot, u_o)$ has an infinite number of zeros for any choice of $u_o$. Each of these zeros represents a phase conjugate to $u_o$. There are an infinite number of such phases in both the past and future of every observer. The discussion in Sect. \ref{Sect:NumImages} therefore implies that under generic conditions, an infinite number of images appear for almost every source. Explicitly, all conjugate phases are given by
\begin{equation}
\tau_n(u_o) = u_o +  n\pi h_1^{ -\frac{1}{2} }, 
\label{Tau}
\end{equation}
where $n$ is any nonzero integer. It is evident from \eqref{IDef} that all of these phases have multiplicity 1. For any $n<0$ and any observation point $p_o$ with $u(p_o) = u_o$, exactly one image of each source is visible as that source appeared in $\mathcal{N}_n(u_o)$. This corresponds to the region between $u = u_o + n \pi h_1^{- \frac{1}{2} }$ and $u = u_o + (n-1) \pi h_1^{- \frac{1}{2} }$. Note that $\det \bB (\cdot, u_o)$ switches sign on each pass through a conjugate phase. The parity of an image emitted from $\mathcal{N}_n(u_o)$ is therefore opposite to the parity of an image emitted from $\mathcal{N}_{n-1}(u_o)$.

Specializing to cases where the source is a geodesic and the observer is instantaneously comoving with that source on $S_{u_o}$, some configurations lead to redshifts and others to blueshifts. Using \eqref{RedshiftGeodesic}, 
\begin{equation}
\frac{ \omega_o}{ \omega_e}  = 1 + \frac{ \delta \bx_o^\intercal \bH \delta \bx_o }{ \kappa_s + |\bB ^{-\intercal} \delta \bx_o|^2}. 
\label{RedshiftGravSym}
\end{equation}
An image is therefore redshifted if and only if 
\begin{equation}
	\delta \bx_o^\intercal \bH \delta \bx_o = - h_1 ( \delta x^1_o)^2 + |h_2| (\delta x^2_o )^2 <0.
\end{equation}
It is blueshifted when $\delta \bx_o^\intercal \bH \delta \bx_o > 0$. There is no frequency shift at all in cases where
\begin{equation}
	(  \delta x_o^1 / \delta x_o^2  )^2 = | h_2/h_1|.
\end{equation}
The direction of the frequency shift clearly depends only on the instantaneous orientation $\delta \bx_o/ | \delta \bx_o|$ of the source and the observer on $S_{u_o}$. In particular, it does not depend on which image is considered. All images of a particular source experience the same type of frequency shift.

Emission times $u_e$ for an observer comoving with a geodesic source may be found by solving \eqref{SigGeo}. For gravity-dominated symmetric waves, this equation reduces to
\begin{eqnarray}
\kappa_s (u_o-u_e) &=& 2 ( \dot{\bg }_o \cdot \delta \bx_o - \delta v_o)+ h_1^{ \frac{1}{2} } (\delta x_o^1)^2 \cot h_1^{ \frac{1}{2} } (u_o-u_e) 
\nonumber
\\
&& ~ + |h_2|^{ \frac{1}{2} } (\delta x^2_o)^2 \coth |h_2|^{ \frac{1}{2} } (u_o-u_e).
\end{eqnarray}
If $-n \gg 1$, it is evident that the image from $\mathcal{N}_n (u_o)$ must satisfy $\cot h_1^{ \frac{1}{2} } (u_o-u_e) \gg 1$. Images from the distant past are therefore emitted at phases $u_e$ very nearly conjugate to $u_o$: \begin{equation}
	u_e \approx \tau_n(u_o) - \frac{ h_1^{ \frac{1}{2} } (\delta x^1_o)^2 }{ |n| \pi \kappa_s  }.
\label{UeGravSym}
\end{equation}
Substituting this relation into \eqref{PsiDef} and \eqref{RedshiftGravSym} shows that very old images cluster near $\ell^a$ on the observer's sky and experience increasingly-negligible frequency shifts:
\begin{equation}
	\psi \propto |n|^{-1}, \qquad |\omega_o/\omega_e-1| \propto |n|^{-2}.
\end{equation}
Old images of slightly extended sources are also highly distorted and demagnified. Their angular diameter and luminosity distances both scale like 
\begin{equation}
	d_\mathrm{ang} \sim d_\mathrm{lum} \propto |n|^{\frac{3}{2}} \exp \left( \frac{1}{2} \sqrt{ | h_2/h_1 | } |n| \pi \right).
\end{equation}	
Gravity-dominated symmetric waves therefore produce an infinite number of exponentially dimming images for almost every source.

\subsection{Matter-dominated symmetric waves}

Matter-dominated symmetric waves satisfying $h_2>0$ act somewhat differently than gravity-dominated waves. In these cases,
\begin{eqnarray}
\bA (u_s,u_o) = \left( \begin{array}{cc}
\cos  h_1^{ \frac{1}{2} } (u_s-u_o)	&	0	\\
0	&	\cos h_2^{ \frac{1}{2} } (u_s-u_o)
\end{array} \right),
\\
\bB (u_s,u_o) = \left( \begin{array}{cc}
h_1^{-\frac{1}{2} } \sin h_1^{ \frac{1}{2} } (u_s-u_o)	&	0	\\
0	&	h_2^{ -\frac{1}{2} } \sin h_2^{ \frac{1}{2} } (u_s-u_o)
\end{array} \right).
\label{BEMWave}
\end{eqnarray}
Phases conjugate to $u_o$ occur at $u_o + n \pi h_1^{- \frac{1}{2} }$ and at $u_o + n' \pi h_2^{- \frac{1}{2} }$, where $n,n'$ are any nonzero integers. If $\sqrt{h_1/h_2}$ is an irrational number, these two families of phases are distinct. Each conjugate pair then has multiplicity 1. If $\sqrt{h_1/h_2}$ is rational, some conjugate pairs have multiplicity 2. In the conformally-flat case where $h_1=h_2$, all conjugate phases have multiplicity 2. In every other case where $\sqrt{h_1/h_2}$ is rational, an infinite number of conjugate phases occur with each multiplicity. Regardless of $h_2$, an infinite number of images are formed for almost every source.

Now consider a luminous source moving on a geodesic. If the source and observer are instantaneously comoving in the sense of \eqref{Comoving2} and \eqref{Comoving3}, frequency shifts associated with each image are given by \eqref{RedshiftGeodesic}. Since \begin{equation}
\bB ^{-1} (\bA \bA ^\intercal - \bm{\delta} ) \bB ^{-\intercal} = \bH
\end{equation}
is negative-definite in this case, all images are redshifted. 

If $-n \gg 1$, an image originating from $\mathcal{N}_n (u_o)$ must be emitted just before the source intersects $S_{\tau_n(u_o)}$. All such images cluster towards $\ell^a$ in the observer's sky and have negligible frequency shifts. Images emitted near conjugate hyperplanes with multiplicity 1 are highly distorted. Images emitted near conjugate hyperplanes with multiplicity 2 are not significantly distorted at all. In both cases, however, older images are dimmer (although the rate at which this occurs is much slower than for gravity-dominated symmetric waves). 

Regardless of the sign of $h_2$, the oldest images formed by symmetric plane wave spacetimes depend on the spacetime structure at arbitrarily large transverse distances. If the metric is modified so that the wave decays at large distances, only a finite number of images discussed here would be unaffected. The oldest images found in pure symmetric waves likely do not appear at all in perturbed symmetric waves.

\section{Sandwich waves}
\label{Sect:Sandwaves}

Symmetric plane waves are mathematically simple, but are not reasonable models for gravitational radiation emitted from compact sources. More interesting are waves where $\bH$ is nonzero only for a finite time: Sandwich waves. Suppose, in particular, that there exists some $u_+ >0$ such that $\bH (u)=0$ for all $u \notin [0,u_+]$. It follows from \eqref{Riemann} that such spacetimes are locally flat whenever $u<0$ or $u>u_+$. The curved region containing the gravitational wave is effectively sandwiched between the two null hyperplanes $S_{0}$ and $S_{u_+}$. Every timelike curve eventually passes entirely through such a wave.

Before an observer interacts with the wave, spacetime is flat and optics is trivial. The case $u_o > u_+$ where an observer has already passed through the wave is more interesting. In this case, $\bA $ and $\bB $ reduce to their flat space forms 
\begin{equation}
	\bA (u_s,u_o) = \bm{\delta}, \qquad \bB (u_s, u_o) = (u_s-u_o) \bm{\delta}
\end{equation}
when $u_s > u_+$. The forms of $\bA $ and $\bB $ inside the wave depend on the details of $\bH $, and will not be discussed here. If $u_s < 0$, however, there always exist four constant $2 \times 2$ matrices $\bm{\alpha}$, $\dot{\bm{\alpha}}$, $\bm{\beta}$ and $\dot{\bm{\beta}}$ such that
\begin{equation}
	\bB (u_s, u_o) = (\bm{\alpha} + \dot{\bm{\alpha}} u_o) +( \bm{\beta} + \dot{\bm{\beta}} u_o) u_s.
	\label{BSandwich}
\end{equation}
Note that the dots on $\dot{\bm{\alpha}}$ and $\dot{\bm{\beta}}$ do not refer to derivatives in this case. They are only used as a labelling device. It follows from \eqref{PrimedD} that $\bA (u_s,u_o)$ is independent of $u_o$. Moreover,
\begin{equation}
	\bA (u_s, u_o) = - \dot{\bm{\alpha}} - \dot{\bm{\beta}} u_s.
\end{equation}
If there were no wave at all, $\bm{\alpha} = \dot{\bm{\beta}}= 0$ and $-\dot{\bm{\alpha}} = \bm{\beta} = \bm{\delta}$.

In general, $\dot{\bm{\alpha}}$ and $\dot{\bm{\beta}}$ have a simple physical interpretation. If two geodesics are comoving and have a transverse separation $\delta \bx_o$ when $u>u_+$, it follows from \eqref{SpatGeo} that the transverse separation between these geodesics is $-\dot{\bm{\alpha}} \delta \bx_o$ immediately before they interact with the wave at $u=0$. Similarly, the relative transverse velocity of these geodesics is $- \dot{\bm{\beta}} \delta \bx_o$ when $u<0$. 

Interpretations for $\bm{\alpha}$ and $\bm{\beta}$ are somewhat less direct. Consider two geodesics which intersect at some time $u_o > u_+$ after the wave has passed, but which have a relative transverse velocity $\delta \dot{\bx}_o$ at $u=u_o$. The difference in transverse velocities between these two geodesics is then $(\bm{\beta} + u_o \dot{\bm{\beta}} ) \delta \dot{\bx}_o$ when $u<0$. Similarly, the difference in the transverse positions of these geodesics is $(\bm{\alpha} + u_o \dot{\bm{\alpha}} ) \delta \dot{\bx}_o$ at $u=0$. It follows that $\bm{\alpha}$ controls shifts in position that are independent of the time $u_o$ at which the two geodesics cross each other.

In principle, $\bm{\alpha}$, $\dot{\bm{\alpha}}$, $\bm{\beta}$, $\dot{\bm{\beta}}$ may all be found by solving \eqref{EDef} if $\bH $ is known. At first glance, this would appear to imply that $4 \cdot 4=16$ numbers are required to describe observations through a sandwich wave. The actual number of required parameters is somewhat less than this. First note that the Wronskian identity \eqref{Wronskian} implies that $\dot{\bm{\alpha}}^\intercal \dot{\bm{\beta}}$ is a symmetric matrix. It also implies that
\begin{equation}
	\dot{\bm{\beta}}^\intercal \bm{\alpha} - \dot{\bm{\alpha}}^\intercal \bm{\beta} = \bm{\delta}.
	\label{WronskSand}
\end{equation}
Further simplifications arise by recalling from \eqref{SymMatrices} that $\bB \bA ^\intercal$ and $\bB ^\intercal \partial_{(1)} \bB $ are symmetric. It follows that
\begin{equation}
	\bm{\alpha} \dot{\bm{\alpha}}^\intercal, \qquad \bm{\beta} \dot{\bm{\beta}}^\intercal, \qquad \bm{\alpha}^\intercal \bm{\beta} 
\end{equation} 
are symmetric as well. These expressions are completely general, and hold for any sandwich wave. They significantly constrain the number of independent parameters needed to specify $\bA $ and $\bB $. Equivalently, they limit the number of parameters that must be extracted from $\bH $. 

It follows from the arguments of Sect. \ref{Sect:NumImages} that the number of images of a generic source observable in any plane wave spacetime is governed by the number of hypersurfaces conjugate to the $u=\mbox{constant}$ hypersurface $S_{u_o}$ containing the observation event $p_o$. Continuing to assume that $u_o > u_+$, all phases conjugate to $u_o$ must be smaller than $u_+$. It follows from \eqref{ZeroDet} and \eqref{BSandwich} that conjugate phases occurring before the wave may be found by solving
\begin{equation}
	\det [ (\bm{\alpha} + \dot{\bm{\alpha}} u_o) +( \bm{\beta} + \dot{\bm{\beta}} u_o) \tau ] = 0
	\label{ConjSand}
\end{equation}
for all $\tau<0$. This equation is quadratic, so at most two solutions exist. An observer ahead of the wave may therefore see at most three images of a source as it appeared behind the wave. There may also be at most one image of a source as it appeared ahead of the wave. In principle, any number of images may arise from inside the wave [where \eqref{BSandwich} is not valid] if $\bH$ is sufficiently large.

Conjugate phases found by solving \eqref{ConjSand} clearly depend on the observation time $u_o$. Less obviously, the \textit{number} of conjugate phases can also depend on $u_o$. For an observer moving on a timelike worldline (where $u_o$ increases monotonically), new conjugate phases -- and therefore new images -- sometimes appear at discrete times.  These images correspond to observation times where \eqref{ConjSand} momentarily degenerates to a linear equation. New images can therefore arise when $u_o = \bar{u}_o$ and 
\begin{equation}
	\det ( \bm{\beta} + \dot{\bm{\beta}} \bar{u}_o ) = 0.
	\label{NewTau}
\end{equation}
If $\det \dot{\bm{\beta}} \neq 0$, the two solutions to this equation are
\begin{equation}
	\fl \qquad  \bar{u}_o = \frac{ - [ \Tr \bm{\beta}  \Tr \dot{\bm{\beta}} - \Tr ( \bm{\beta} \dot{\bm{\beta}})] \pm \sqrt{  [ \Tr \bm{\beta}  \Tr \dot{\bm{\beta}} - \Tr ( \bm{\beta} \dot{\bm{\beta}})]^2 - 4 \det \bm{\beta} \det \dot{\bm{\beta}} } }{ 2 \det \dot{\bm{\beta}} }. 
\end{equation}
Only solutions satisfying $\bar{u}_o >u_+>0$ are physically relevant. When a conjugate phase of this type first appears, it satisfies
\begin{equation}
	\lim_{u_o \rightarrow \bar{u}_o^+} \tau(u_o) = - \infty.
\end{equation}
The associated image therefore provides a picture of the infinitely distant past. Furthermore, an infinite amount of the source's history appears to the observer within a finite amount of proper time. This implies that new images are highly blueshifted. 

No matter how long an observer waits, no conjugate phase can exceed $u_+$. The emission time for an associated image might therefore be expected to tend towards a constant value as $u_o \rightarrow \infty$. This means that a very large amount of proper time at the observer corresponds to only a small amount of proper time at a source. Images which appear suddenly and are initially highly blueshifted become highly redshifted at late times. 

It is unclear precisely what these types of images imply. To the extent that geometric optics remains valid, all observers passing through $S_{\bar{u}_o}$ momentarily see almost the entire universe appear infinitely blueshifted as it was in the infinitely distant past. Furthermore, \eqref{AngleRedshift} implies that all of the universe is briefly compressed into a single point on each observer's sky. Of course, such phenomena lie outside the domain of geometric optics. They may even lie outside of the realm of test fields propagating on a fixed background spacetime. Extreme focusing events like these might indicate instabilities inherent in the plane wave geometry itself. It should, however, be noted that all of the infinities just alluded to are likely to have finite cutoffs in ``realistic'' plane waves which decay at large transverse distances.

\subsection*{Weak wavepackets}
\label{Sect:WeakWave}

One important class of sandwich waves are those that are very weak and last only for a short time. In these cases, $\bm{\alpha}$, $\dot{\bm{\alpha}}$, $\bm{\beta}$, and $\dot{\bm{\beta}}$ may be expanded as integrals involving successively higher powers of $\bH $. To lowest order in such a scheme, $\bA $ and $\bB $ are approximately unaffected by the wave while inside of it. Assuming that $-u_s, u_o \gg u_+$, the first corrections to this assumption are
\begin{eqnarray}
	\bA (u_s, u_o) \approx \bm{\delta} - u_s \int_{0}^{u_+} \rmd w \mathbf{H}(w),
	\label{ASand}
	\\
	\bB (u_s,u_o) \approx (u_s-u_o) \bm{\delta}  + u_s u_o \int_0^{u_+} \rmd w  \bH(w).
	\label{BSand}
\end{eqnarray}
This approximation is consistent with \eqref{Wronskian} [and therefore \eqref{WronskSand} as well]. In terms of the matrices appearing in \eqref{BSandwich}, $\bm{\alpha} \approx 0$, $-\dot{\bm{\alpha}} \approx \bm{\beta} \approx \bm{\delta}$, and
\begin{equation}
	\dot{\bm{\beta}} \approx \int_{0}^{u_+} \rmd w \mathbf{H}(w) .
\end{equation}
Note that Eqs. \eqref{ASand} and \eqref{BSand} should be applied with care if $\bH $ involves many oscillations of an approximately periodic function. In these cases, the integral of $\bH $ can be very nearly zero. Terms nonlinear in $\bH $ might then be significant. 

Assuming that \eqref{ASand} and \eqref{BSand} are indeed adequate approximations for $\bA $ and $\bB $, a coordinate rotation may always be used to diagonalize $\dot{\bm{\beta}}$. There then exist two constants $\mathcal{H}_1$ and $\mathcal{H}_2$ such that 
\begin{equation}
\dot{\bm{\beta}} = \left( \begin{array}{cc}
			-\mathcal{H}_1  &	0	\\
			0	&	 - \mathcal{H}_2
	\end{array} \right).
\label{HInt}
\end{equation}
In this sense, all sufficiently short gravitational plane waves act as though they are linearly polarized [so $\beta=\chi=0$ in \eqref{Distortion}]. In terms of the individual wavefunctions appearing in \eqref{HGen},
\begin{eqnarray}
	\mathcal{H}_1 &=& \int_0^{u_+} \rmd w [h_\| (w) + h_+(w)] ,
	\\
	\mathcal{H}_2 &=& \int_0^{u_+} \rmd w [h_\|(w) - h_+(w)].
\end{eqnarray}
The transverse coordinates $x^i$ have also been chosen such that
\begin{equation}
	\int_0^{u_+} \rmd w h_\times(w) =0.
\end{equation}

If a wave satisfies the vacuum Einstein equation, $h_\| = 0$ and $\mathcal{H}_1 = - \mathcal{H}_2$. More generally, it follows from the weak energy condition that
\begin{equation}
	\mathcal{H}_1 + \mathcal{H}_2 \geq 0.
	\label{HInequality}
\end{equation}
Now assume that the integral of $h_+$ is non-negative, which entails only a minimal loss of generality. Then, 
\begin{equation}
	\mathcal{H}_1 > 0, \qquad \mathcal{H}_1 \geq |\mathcal{H}_2|.
\end{equation} 
We say that a wave is gravity-dominated if $\mathcal{H}_2 < 0$ and matter-dominated if $\mathcal{H}_2 >0$. These definitions are closely analogous to those used to classify symmetric plane waves in Sect. \ref{Sect:SymWaves}. There, a wave was said to be gravity- or matter-dominated depending on the sign of the constant $h_2 = h_\| - h_+$ appearing in \eqref{SymWave}.

For weak gravity-dominated wavepackets, there can be at most one phase conjugate to an observer satisfying $u_o>u_+$. If this exists, it evident from \eqref{BSand} that
\begin{equation}
	\tau_{-1} (u_o) = - \frac{ u_o }{ \mathcal{H}_1 u_o -1} 
	\label{Tau1}
\end{equation}
is conjugate to $u_o$ with multiplicity 1. This equation is valid only if $\tau_{-1}(u_o)<0$. A conjugate phase therefore exists only for observers satisfying
\begin{equation}
	\mathcal{H}_1  u_o >1 .
	\label{Observer1}
\end{equation}
Note that $\mathcal{H}_1 \bar{u}_o = 1$ is the unique physically-relevant solution to \eqref{NewTau} in the gravity-dominated case.

Waves that are matter-dominated (so $\mathcal{H}_2 >0$) also admit the conjugate phase \eqref{Tau1} when $u_o$ satisfies \eqref{Observer1}. In the conformally-flat case where $\mathcal{H}_1 = \mathcal{H}_2$, this is the only conjugate phase. Unlike in the gravity-dominated case, the multiplicity of $\tau_{-1}$ is equal to 2 for conformally-flat waves. In all other matter-dominated cases, $\tau_{-1}$ has multiplicity 1 and a second conjugate phase is admitted (also with multiplicity 1)
for all observers satisfying 
\begin{equation}
\mathcal{H}_2 u_o > 1.
\label{Observer2}
\end{equation}
This occurs at
\begin{equation}
	\tau_{-2} (u_o) = - \frac{ u_o }{ \mathcal{H}_2 u_o -1}.
	\label{Tau2}
\end{equation}
Note that \eqref{Observer2} is a more stringent condition than \eqref{Observer1}. As implied by the notation, $\tau_{-2}(u_o) < \tau_{-1}(u_o)$.

\begin{figure}[t]
	\centering
	\begin{subfigure}[b]{.46\linewidth}
		\centering
		\includegraphics[width= \linewidth]{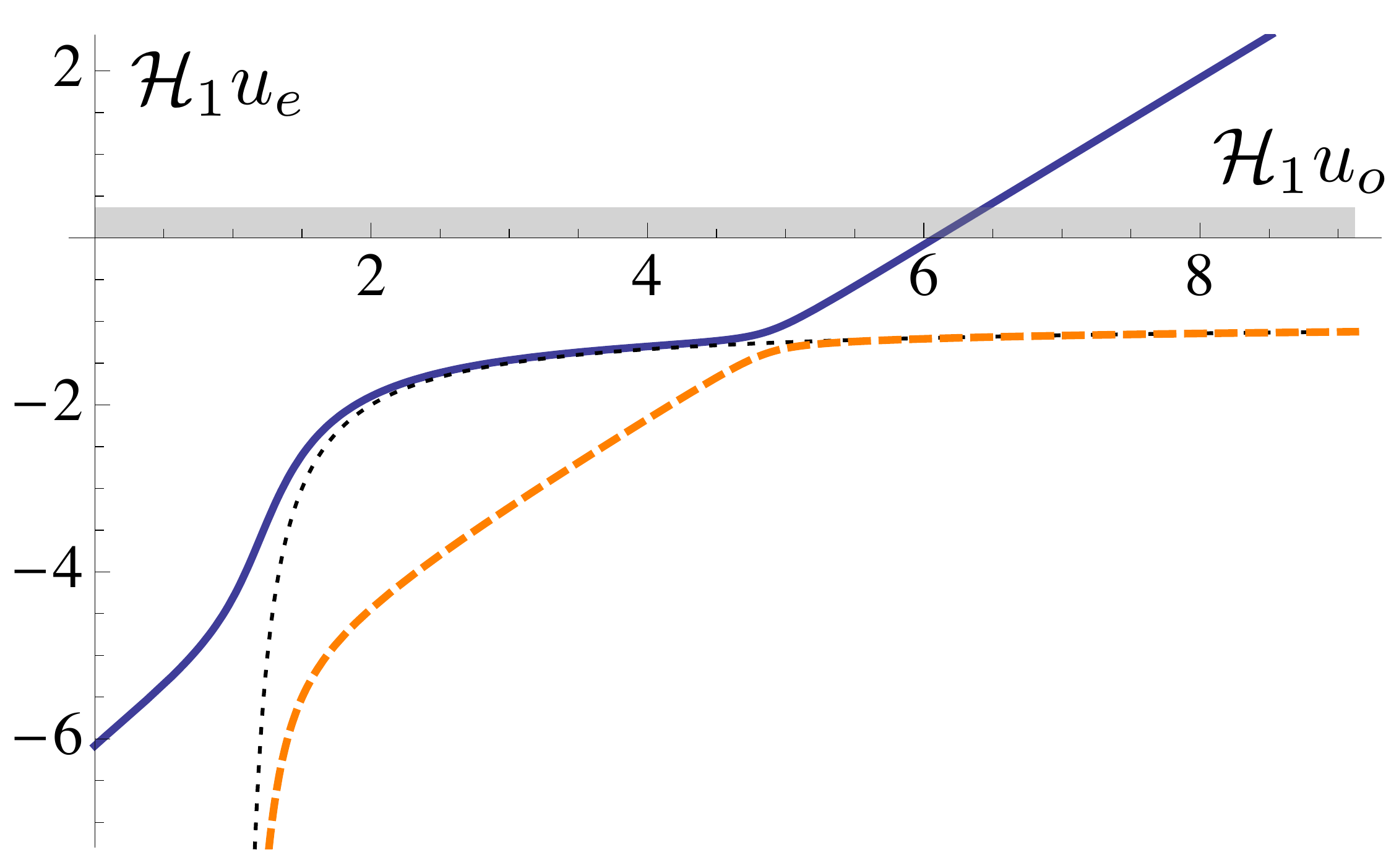}
	\end{subfigure}
	~
	\begin{subfigure}[b]{.46\linewidth}
		\centering
		\includegraphics[width= \linewidth]{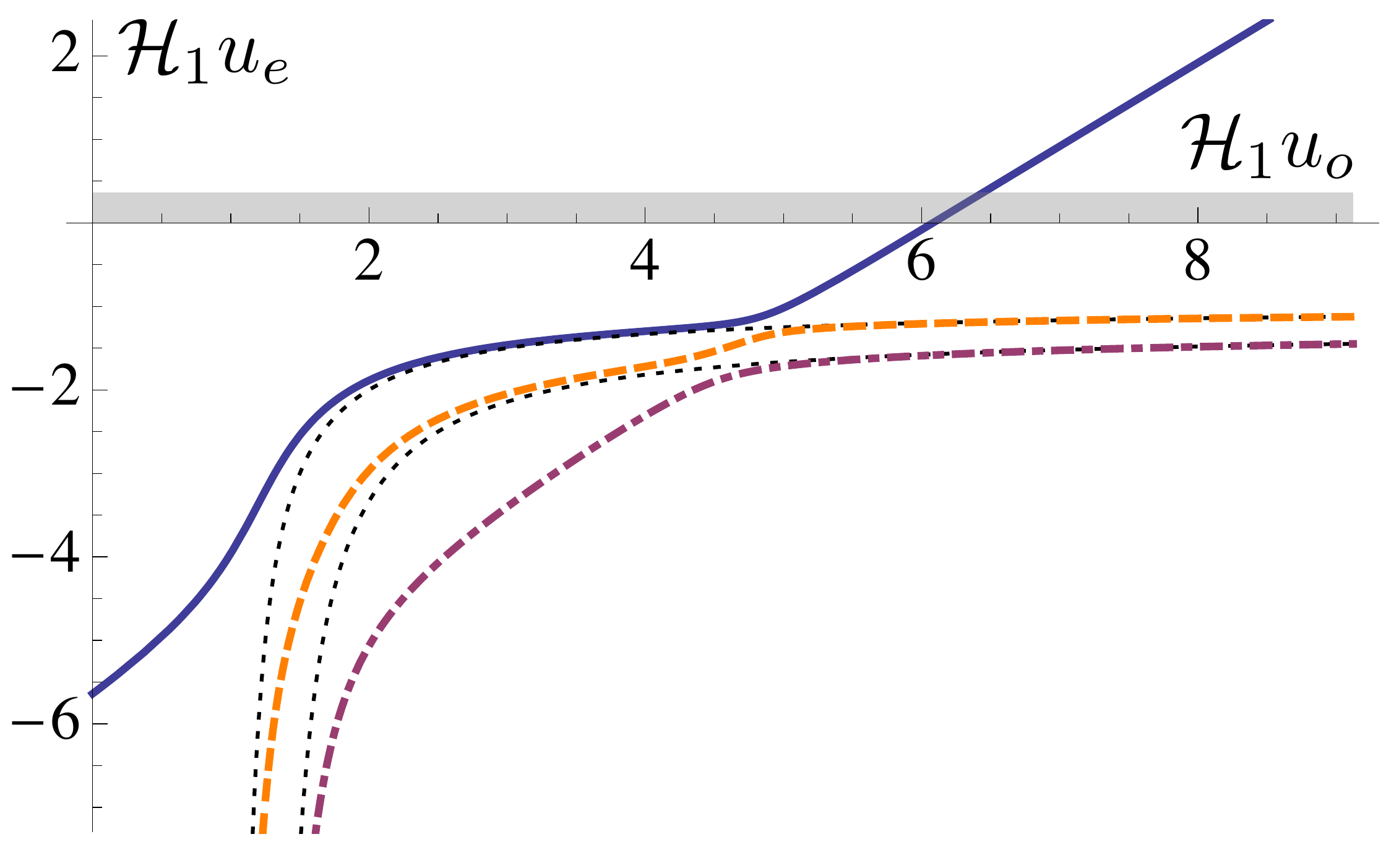}
	\end{subfigure}
	
	\caption{Example emission versus observation times in the presence of weak wavepackets. The left figure assumes a vacuum wave, while the right figure is a matter-dominated wave with $\mathcal{H}_2 = 4 \mathcal{H}_1/5$. Dotted lines represent $\tau_{-1}(u_o)$ and $\tau_{-2}(u_o)$. The wave's location is indicated schematically by a thin grey rectangle. The source and observer are placed on geodesics assumed to be comoving after the wave has passed. In both cases, $\kappa_s = 1$, $\mathcal{H}_1 \delta x_o^1= \mathcal{H}_1  \delta x_o^2 = 1/2$, and $\mathcal{H}_1 (\dot{\bg }_o \cdot \delta \bx_o - \delta v_o ) = 3$.}
	\label{Fig:EmissionSand}	
\end{figure}

\begin{figure}[b]
	\centering
	
	\begin{subfigure}{.47\linewidth}
		\centering
		\includegraphics[width= \linewidth]{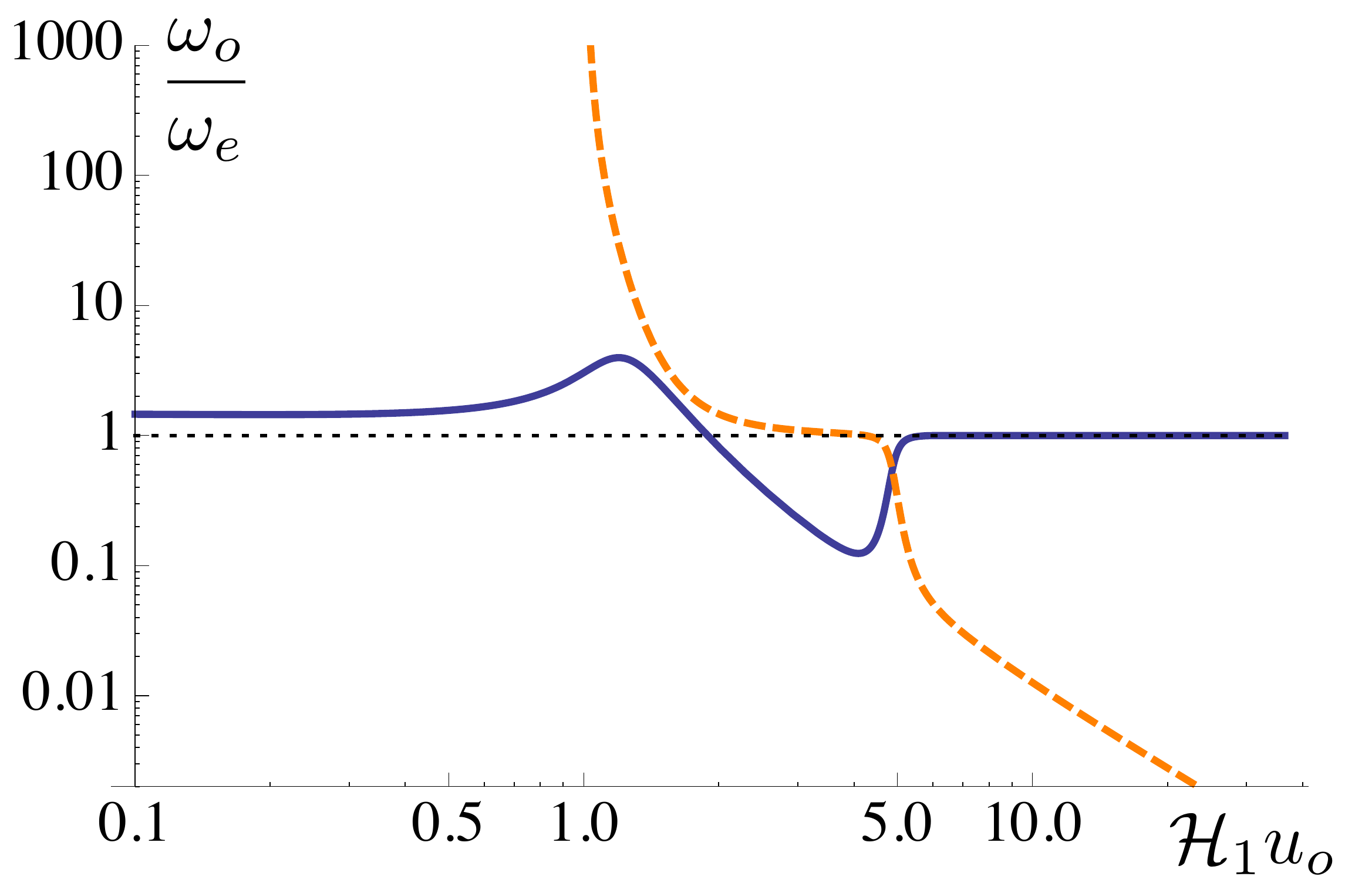}
	\end{subfigure}
	
	\caption{Frequency shifts for a vacuum wavepacket. The parameters and dashing used here are the same as those in the left panel of Fig. \ref{Fig:EmissionSand}. The younger image (solid) experiences a temporary blueshift and then a redshift before settling down to $\omega_o=\omega_e$. The slight initial blueshift as $u_o \rightarrow 0^+$ is due to the source and observer moving towards each other before interacting with the wave. The older image (dashed) initially appears with infinite blueshift. It then suffers an ever-increasing redshift.}
	
	\label{Fig:Redshift}
\end{figure}

Consider a point source moving on a timelike worldline $\Gamma$ in a weak sandwich wave spacetime. Such a source appears differently when observed at different times. It is clear that when $u_o<0$, exactly one image of $\Gamma$ is viewable under generic conditions. As time passes, the wave eventually passes through the observer. A second image then appears when $u_o  = \mathcal{H}_1^{-1} >u_+$. This image is always emitted before the first. If the wave involves a sufficient amount of Ricci curvature\footnote{Distinguishing between different cases based on the Ricci tensor inside a wave requires that the approximations leading to \eqref{ASand} and \eqref{BSand} be valid. It is possible for, e.g., sufficiently strong vacuum waves to admit two conjugate phases in the region $u<0$.} (from e.g., electromagnetic plane waves) and is not conformally-flat, a third image appears when $u_o = \mathcal{H}_2^{-1} > \mathcal{H}_1^{-1}$. This is emitted before the first two images. All images persist indefinitely once they appear. Sufficiently far in the future, one image is observed of the source as it appeared after interacting with the gravitational wave. All other images predate this interaction. See Fig. \ref{Fig:EmissionSand}.

When the second image first appears at $u_o = \mathcal{H}_1^{-1}$, the new conjugate phase $\tau_{-1}(u_o)$ is divergent. At all later times, it is finite. The same is also true for the emission times associated with the second image. Almost the entire past history of the source is therefore observable within a finite proper time. This implies an infinite blueshift. At late times, $\tau_{-1} \rightarrow -\mathcal{H}_1^{-1}$. The observed evolution of the source via the second image effectively freezes as $u_e$ asymptotes to $- \mathcal{H}_1^{-1}$ (which predates the source's interaction with the wave). Images such as these are highly redshifted, as indicated in Fig. \ref{Fig:Redshift}. Note that a similar transition from infinite blueshift to infinite redshift also applies to the third image if it exists. 

\begin{figure}[t]
	\centering
	
	\includegraphics[width=.47\linewidth]{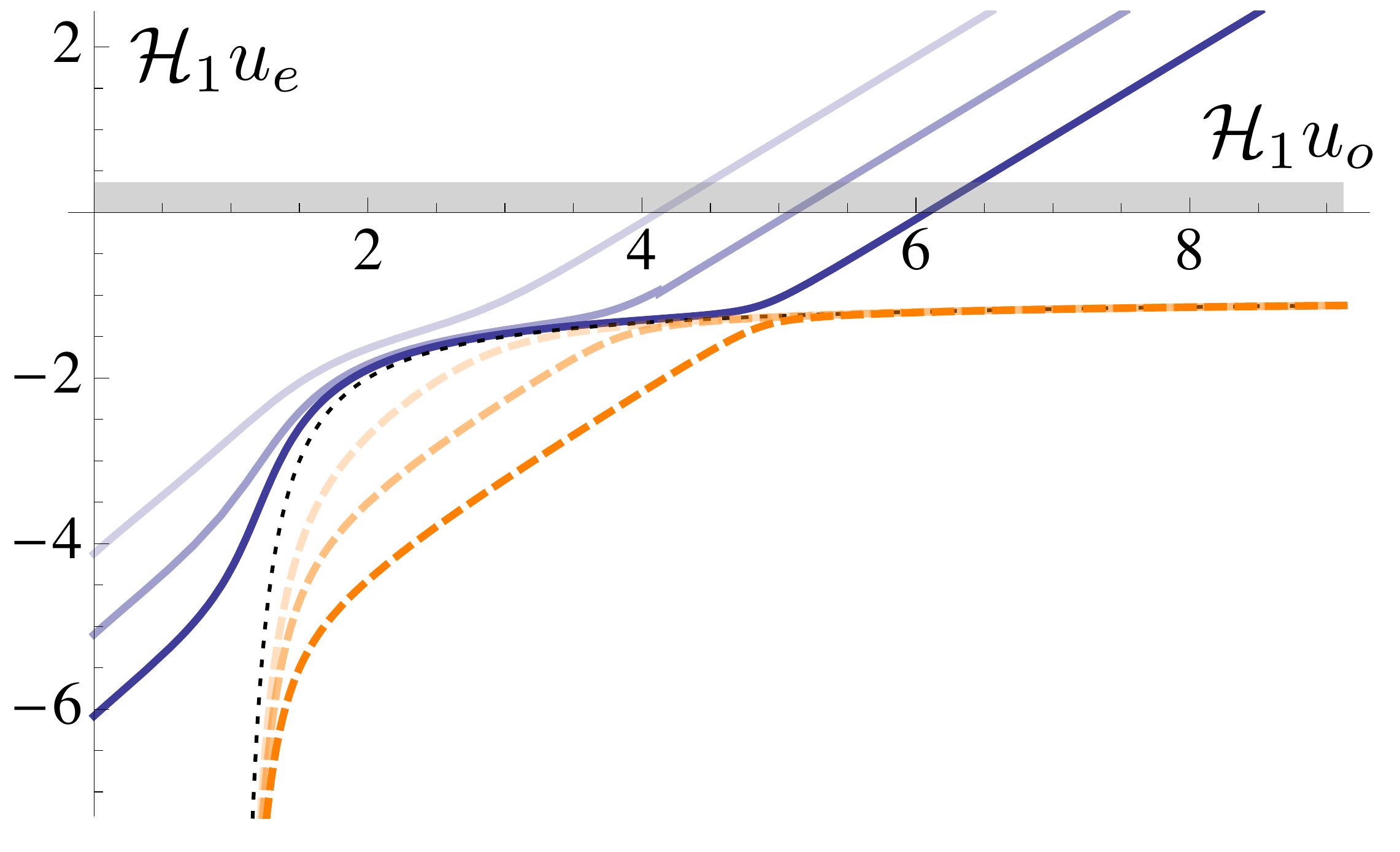}
	
	\caption{Bending of emission curves for a vacuum wavepacket. All parameters are the same as in the left panel of Fig. \ref{Fig:EmissionSand} except that three different choices are made for the value of $\delta v_o$. Curves for both images are shifted to the right as $\delta v_o$ is decreased. The younger (solid) curves are almost unaffected by the wave if $\delta v_o$ is sufficiently large. For smaller values of $\delta v_o$, both curves are strongly bent by the constraint that they can't pass through the dotted curve representing $\tau_{-1}$.}
	
	\label{Fig:ChangingV}
\end{figure}

\begin{figure}[t]
	\centering
	\begin{subfigure}[b]{.46\linewidth}
		\centering
		\includegraphics[width= \linewidth]{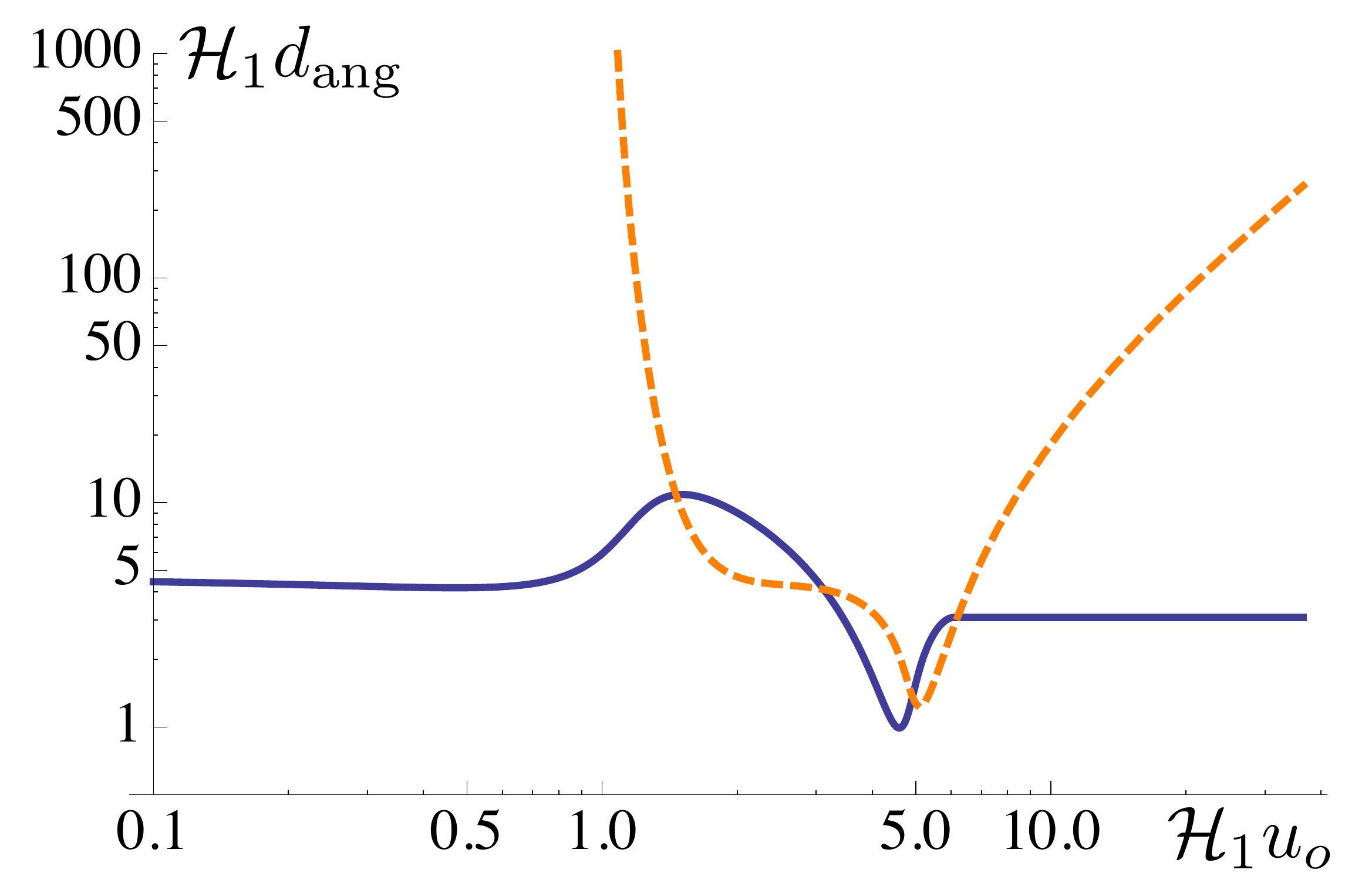}
	\end{subfigure}
	~
	\begin{subfigure}[b]{.46\linewidth}
		\centering
		\includegraphics[width= \linewidth]{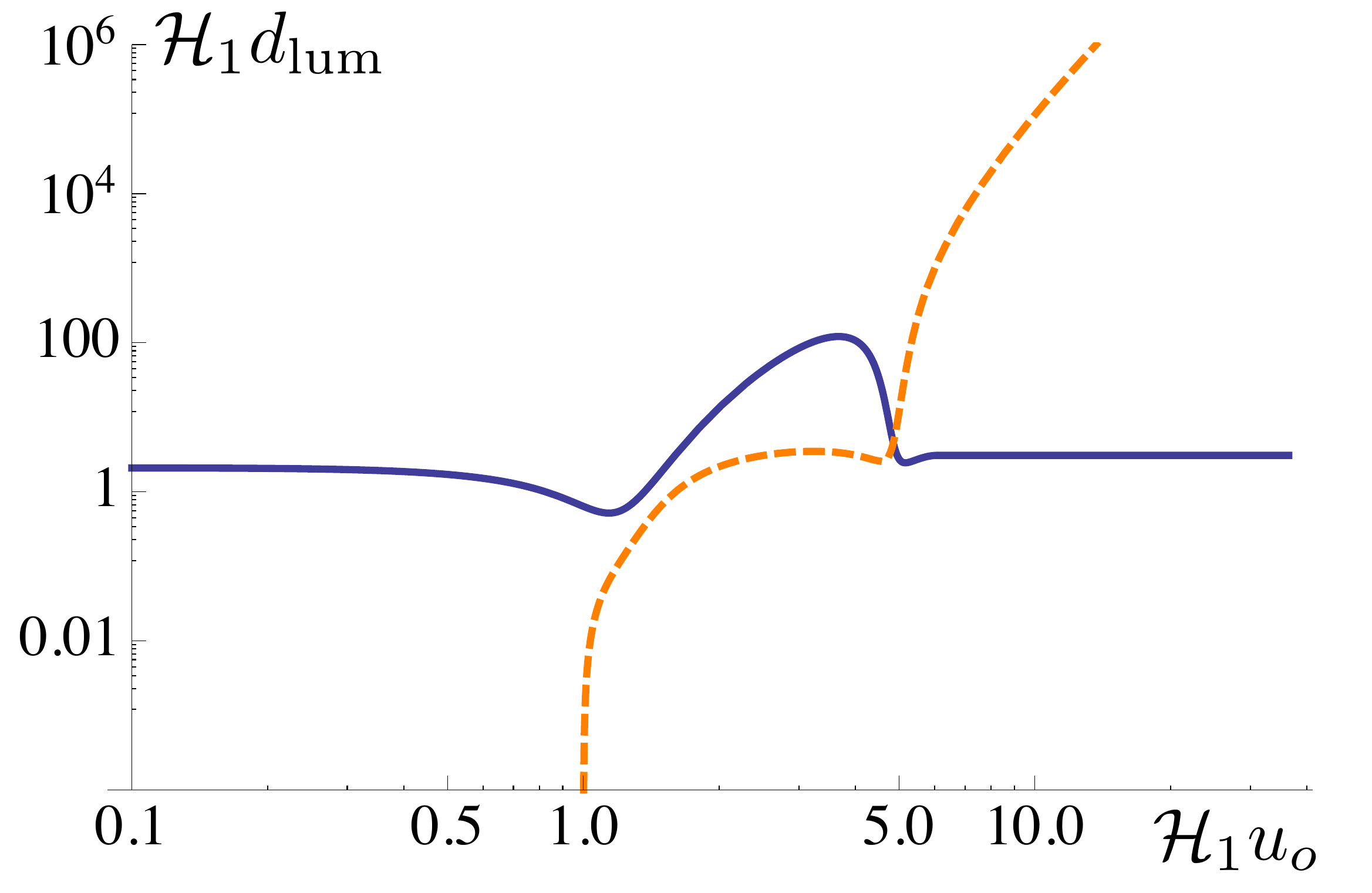}
	\end{subfigure}
	
	\caption{Angular diameter and luminosity distances for a vacuum wavepacket. All parameters are the same as those used in the left panel of Fig. \ref{Fig:EmissionSand}.}
	
	\label{Fig:Luminosity}
\end{figure}

Another qualitative feature of the emission times plotted in Fig. \ref{Fig:EmissionSand} is that there is a sense in which pairs of images can ``switch roles.'' Consider, e.g., the left panel of that figure. At late times, the solid curve (corresponding to the younger image) is perfectly linear. Indeed, it remains very nearly linear until $\mathcal{H}_1 u_o \approx 5$. A rapid transition then occurs where the dashed curve effectively takes over this linear behavior while the solid curve strongly deviates from it. In a sense, the two images reverse their roles. This phenomenon occurs one more time (somewhat less sharply) around $\mathcal{H}_1 u_o \approx 1$ when the second image first appears. It arises essentially because $u_e> \tau_{-1}$ for the younger image and $u_e < \tau_{-1}$ for the older image. These constraints can cause emission curves to bend sharply -- with large accompanying frequency shifts -- in order to avoid intersecting $\tau_{-1}$. Whether or not this occurs depends on whether the ``average'' linear increase of $u_e$ ever comes near $\tau_{-1}$. If it does, this role switching occurs. If not, the younger image is barely affected by the gravitational wave at all. This is illustrated in Fig. \ref{Fig:ChangingV}, where emission curves for several sources are plotted simultaneously.

\begin{figure}[b]
	\centering
	\begin{subfigure}[b]{.46\linewidth}
		\centering
		\includegraphics[width= \linewidth]{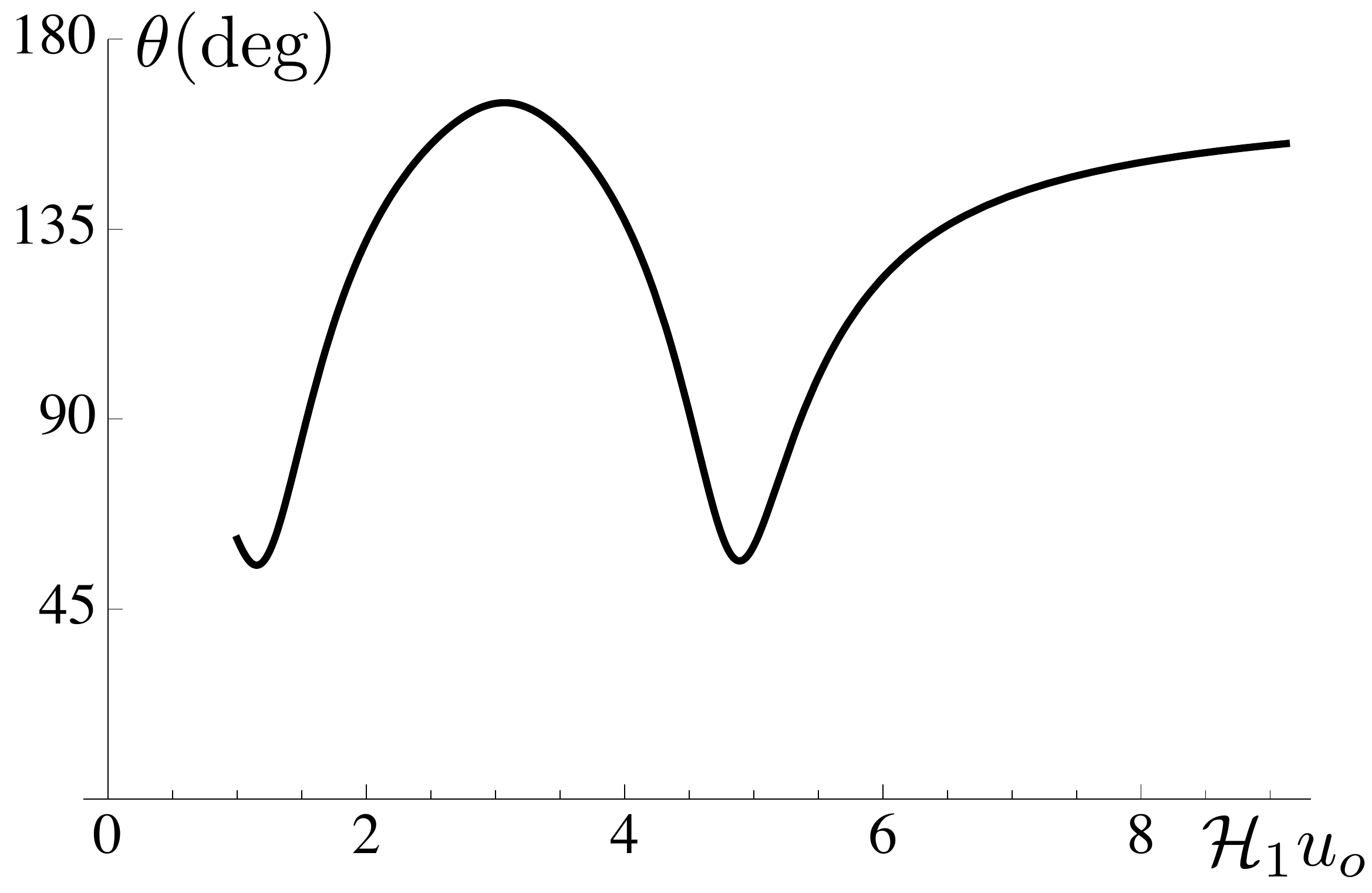}
	\end{subfigure}
	~
	\begin{subfigure}[b]{.46\linewidth}
		\centering
		\includegraphics[width= \linewidth]{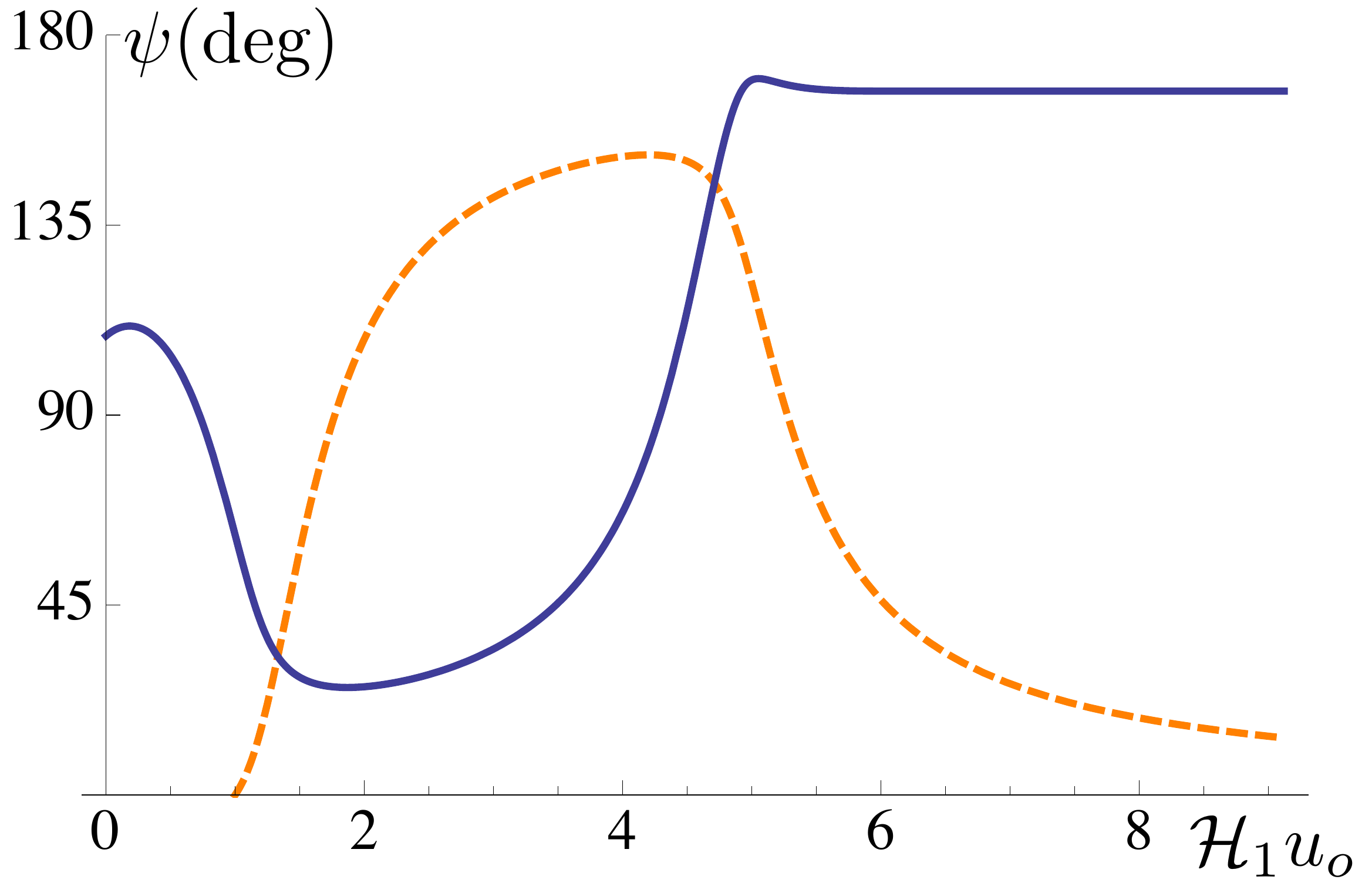}
	\end{subfigure}
	
	\caption{Angles (in degrees) for images produced by a vacuum wavepacket with the same parameters as those used in the left panel of Fig. \ref{Fig:EmissionSand}. $\theta$ measures the angle between both images, while $\psi$ measures the angle between each image and $\ell^a$ on an observer's sky.}
	
	\label{Fig:Angles}
\end{figure}

To summarize, images which appear at discrete times briefly appear as bright, highly blueshifted ``flashes.'' Indeed, Fig. \ref{Fig:Luminosity} shows that their luminosity distances go to zero. Simultaneously, the angular diameter distance of each new image tends to infinity. It is implied by \eqref{AngleRedshift} that all highly blueshifted images make must a very small angle $\psi$ with $\ell^a$ on the observer's sky. The second (and third) images therefore appear aligned with the direction of propagation of the gravitational wave when they first appear. This direction could be quite different from the location of the other image(s). As time progresses, all images migrate across the observer's sky as illustrated in Fig. \ref{Fig:Angles}. Different images may remain separated from each other by large angles at all times.

\section{Discussion}

Despite their simplicity, plane gravitational waves behave in qualitatively different ways from lenses associated with quasi-Newtonian mass distributions. As expected from their dynamic nature, plane waves generically shift the observed frequencies of various images. They may also admit images which appear to move, deform, change brightness, and shift color as time progresses. 

More subtle differences relate to the number of images that are produced  of a given source. For example, even numbers of images can appear generically (which has led plane wave spacetimes to be cited \cite{PerlickReview} as well-behaved examples where the odd number theorem \cite{OddNumber, PerlickOddNumber} does not apply). Some plane waves can even produce an infinite number of discrete images. Perhaps most striking of all are the bright flashes shown to be produced by generic sandwich waves in Sect. \ref{Sect:Sandwaves}. These correspond to individual images which appear at discrete times. More typical gravitational lenses can produce new images if a source crosses an observer's caustic. Individual images then split into two (or vice versa). The flashes produced by sandwich waves are quite different. Their appearance does not require that a source pass through an observer's caustic. Such images appear individually from the infinitely distant past. Initially, they are infinitely bright and infinitely blueshifted points of light appearing in the direction of propagation associated with the gravitational wave.

Many of these effects depend at least partially on the idealization that a plane wave extends undiminished to infinitely-large transverse distances. Plane wave spacetimes are not asymptotically flat. Despite being topologically trivial and locally well-behaved, they are not even globally hyperbolic: Null geodesics passing between appropriately-chosen pairs of points can extend to arbitrarily large transverse distances in between those points. It is this property which permits the infinite number of images described in Sect. \ref{Sect:SymWaves} to be produced by symmetric waves. The flashes described in Sect. \ref{Sect:Sandwaves} also depend on the spacetime structure at arbitrarily large distances. This structure likely affects the formation of even numbers of images as well. Indeed, the usual proofs of the odd number theorem require global hyperbolicity, among other assumptions \cite{PerlickReview, OddNumber} (see, however, \cite{PerlickOddNumber} for a more general formulation).

If a spacetime has the geometry of a plane wave only out to some finite transverse distance, all results derived here remain valid if the associated images involve light rays which never extend sufficiently far to interact with any large-distance modifications. The infinite sequence of images formed by a symmetric plane wave would then be expected to become finite for spacetimes which are only approximately plane waves. Calculations involving the oldest images could no longer be trusted in these cases. Similarly, the bright flashes associated with ideal sandwich waves are likely to be somewhat less extreme for waves which decay at infinity. Large brightnesses and large blueshifts can still exist, but these will be cut off at some finite maximum. Such maxima may, however, remain quite large.

It is reasonably clear that modifications of the geometry at large distances can remove some images. Less obviously, these modification can also introduce new images. Consider, for example, the $pp$-wave spacetimes obtained by substituting
\begin{equation}
	H_{ij} (u) x^i x^j \rightarrow H(u,\bx)
\end{equation}
in the metric \eqref{PlaneWaveMetric}. These generalize the plane wave spacetimes. It has been shown that if $H$ grows subquadratically as $|\bx| \rightarrow 0$, the resulting geometries are globally hyperbolic \cite{PlaneWaveCausality}. Moreover, every pair of points is connected by at least one geodesic in these cases (unlike in pure plane wave spacetimes where the growth of $H$ is precisely quadratic). This implies that modifications of the geometry at large transverse distances can introduce new null geodesics even between points at small transverse distances. It would be interesting to explore these effects in more depth to understand precisely how modifications of this sort (or more general ones) alter the lensing properties described here for ideal plane waves. It would also be interesting to better understand what the transient flashes of Sect. \ref{Sect:Sandwaves} imply for waves propagating on plane (or almost-plane) wave spacetimes. This can likely be facilitated by the Green functions derived in \cite{HarteCaustics}.

 \end{document}